\documentclass[prd,aps,floats,floatfix,eqsecnum,nofootinbib]{revtex4}
\usepackage{verbatim,graphicx,amssymb,amsbsy,bm,amsmath,rotating,epsfig}
\usepackage{psfrag}
\usepackage{hyperref}
\usepackage{fontenc}
\usepackage[latin1]{inputenc}
\usepackage{pstricks,pst-node,pst-text,pst-3d}
\usepackage{textcomp}
\newcommand{\be}{\begin{equation}}
\newcommand{\ee}{\end{equation}}
\newcommand{\bea}{\begin{eqnarray}}
\newcommand{\eea}{\end{eqnarray}}

\newcommand{\br}{\ensuremath{\mathbf{r}}}

\begin{document}
\title{Equation of state, universal profiles, scaling
and macroscopic quantum effects in Warm Dark Matter galaxies}
\author{\bf H. J. de Vega $^{(a,b)}$}
\email{devega@lpthe.jussieu.fr} 
\author{\bf N. G. Sanchez $^{(b)}$}
\email{Norma.Sanchez@obspm.fr} 
\affiliation{$^{(a)}$ LPTHE, Universit\'e Pierre et Marie Curie (Paris VI),
Laboratoire Associ\'e au CNRS UMR 7589, Tour 24, 5\`eme. \'etage, 
Boite 126, 4, Place Jussieu, 75252 Paris, Cedex 05, France. \\
$^{(b)}$ Observatoire de Paris,
LERMA. Laboratoire Associ\'e au CNRS UMR 8112.
 \\61, Avenue de l'Observatoire, 75014 Paris, France.}
\date{\today}
\begin{abstract}
The Thomas-Fermi approach to galaxy structure determines selfconsistently and non-linearly
the gravitational potential of the fermionic 
warm dark matter (WDM) particles given their quantum distribution function $ f(E) $.
This semiclassical framework accounts for the quantum nature and high number of DM particles, properly  
describing gravitational bounded and quantum macroscopic systems as neutron stars,
white dwarfs and WDM galaxies. We express the main galaxy magnitudes as the halo radius 
$ r_h $, mass $ M_h $, velocity dispersion and phase space density 
in terms of the surface density which is important to confront to observations. 
From these expressions we {\bf derive} the general equation of 
state for galaxies, i. e., the relation between pressure and density, and provide
its analytic expression. Two regimes clearly show up: (i) Large diluted galaxies for $ M_h \gtrsim 2.3 \;
10^6 \; M_\odot $ and effective temperatures $ T_0 > 0.017 $ K described by 
the classical selfgravitating WDM Boltzman gas with an inhomogeneous perfect gas equation of state,
and (ii) Compact dwarf galaxies for  $ 1.6 \; 10^6 \; M_\odot \gtrsim M_h \gtrsim M_{h,min} \simeq 
3.10 \; 10^4 \; \left(2 \, {\rm keV}/m\right)^{\! \! \frac{16}5} \; M_\odot, 
\; T_0 < 0.011 $ K described by the quantum fermionic WDM regime with a steeper equation of 
state close to the degenerate state. In particular, the $ T_0 = 0 $ degenerate or extreme quantum limit yields 
the most compact and smallest galaxy. All magnitudes in the diluted regime turn to exhibit square 
root of $ M_h $ {\bf scaling} laws and are {\bf universal} functions of $ r/r_h $ 
reflecting the WDM perfect gas behaviour in this regime. These theoretical results contrasted to 
robust and independent sets of galaxy data 
remarkably reproduce the observations. For the small galaxies, $ 10^6 \gtrsim M_h \geq M_{h,min} $,
the equation of state is galaxy mass dependent 
and the density and velocity profiles are not anymore universal,
accounting to the quantum physics of the self-gravitating WDM fermions in the compact regime (near, but not at, 
the degenerate state). It would be extremely interesting to dispose of dwarf galaxy 
observations which could check these quantum effects. 
\end{abstract}
\pacs{95.35.+d, 98.52.-b, 98.52.Wz}
\keywords{Dark Matter, Galaxy structure, Galaxy Density Profiles}
\maketitle
\tableofcontents

\section{INTRODUCTION}

Dark matter (DM) is the main component of galaxies: the fraction of DM over the total
galaxy mass goes from 95\% for large diluted galaxies till 99.99\% for dwarf compact galaxies.
Therefore, DM alone should explain the main structure of galaxies. Baryons should only give
corrections to the pure DM results.

\medskip

Warm Dark Matter (WDM), that is dark matter formed by particles with masses in the keV scale
receives increasing attention today (\cite{highp,highm,cosmf} and references therein).

{\vskip 0.1cm} 

At intermediate scales $ \sim 100 $ kpc, WDM gives the {\bf correct abundance} of substructures 
and therefore WDM solves the cold dark matter (CDM) overabundance of structures at small scales.
\citep{colin,dolgov,theuns,tikho,zav,pap,lov12,lovl,ander}.
For scales larger than $ ~ 100 $ kpc, WDM yields the same results than CDM.
Hence,  WDM agrees with all the observations:
small scale as well as large scale structure observations and CMB anisotropy observations.

{\vskip 0.1cm} 

Astronomical observations show that the DM galaxy density profiles are {\bf cored} 
till scales below the kpc \cite{obs,gil,wp}.
On the other hand, $N$-body CDM simulations exhibit cusped density profiles with a typical $ 1/r $ behaviour
near the galaxy center $ r = 0 $. 
Inside galaxy cores, below  $ \sim 100$ pc, $N$-body classical physics simulations 
do not provide the correct structures for WDM because quantum effects are important in WDM at these scales.
Classical physics $N$-body WDM simulations exhibit cusps or small 
cores with sizes smaller than the observed cores \citep{avi,colin8,vinas,sinz}.
WDM predicts correct structures and cores with the right sizes for small scales (below kpc) 
when the {\bf quantum} nature of the WDM particles is taken into account \cite{newas,astro}.
This approach is {\bf independent} of any WDM particle physics model.

\medskip

We follow here the Thomas-Fermi approach to galaxy structure for self-gravitating 
fermionic WDM \citep{newas,astro}. This approach is especially appropriate to take into account quantum properties
of systems with large number of particles. That is, macroscopic quantum systems
as neutron stars and white dwarfs \citep{ll}.
In this approach, the central quantity to derive is the DM chemical potential $ \mu(\br) $,
which is the free energy per particle. For self-gravitating systems,
the potential $ \mu(\br) $ is proportional to the gravitational potential $ \phi(\br) $,
$ \mu(\br) =  \mu_0 - m \; \phi(\br) $, $ \mu_0 $ being a constant, and 
obeys the {\bf self-consistent} and {\bf nonlinear} Poisson equation
\be\label{poisI}
\nabla^2 \mu(\br) = -4 \; \pi \; g \; G \; m^2 \; 
\int \frac{d^3p}{(2 \, \pi \; \hbar)^3} \; f\left(\frac{p^2}{2 \, m}-\mu(\br)\right) \; .
\ee
Here $ G $ is Newton's gravitational constant,  $ g $ is the number of internal degrees of freedom 
of the DM particle, $ p $ is the DM particle momentum and $ f(E) $ is the
energy distribution function. This is a semiclassical gravitational approach to determine selfconsistently the
gravitational potential of the quantum fermionic WDM given its distribution function $ f(E) $.

\medskip

In the Thomas-Fermi approach, DM dominated galaxies are considered in a stationary state.
This is a realistic situation for the late stages of structure formation
since the free-fall (Jeans) time $ t_{ff} $ for galaxies is much shorter than the age of galaxies.
$ t_{ff} $ is at least one or two orders of magnitude smaller than the age of the galaxy.

\medskip

We consider spherical symmetric configurations where eq.(\ref{poisI}) becomes an
ordinary nonlinear differential equation that determines self-consistently
the chemical potential $ \mu(r) $ and constitutes the Thomas--Fermi approach \citep{newas,astro}.
We choose for the energy distribution function a Fermi--Dirac distribution
$$
f(E) = \frac1{e^{E/T_0} + 1} \; ,
$$
where $ T_0 $ is the characteristic one--particle energy scale. $ T_0 $ plays
the role of an effective temperature scale and
depends on the galaxy mass. The Fermi--Dirac distribution function
is justified in the inner regions of the galaxy,
inside the halo radius where we find that the Thomas--Fermi density
profiles perfectly agree with the observations. 

\medskip

The solutions of the Thomas--Fermi equations (\ref{poisI})
are characterized by the value of the chemical potential at the origin  $ \mu(0) $.
Large positive values of $ \mu(0) $ correspond to dwarf 
compact galaxies (fermions near the quantum degenerate limit),
while large negative values of $ \mu(0) $ yield large and diluted galaxies (classical 
Boltzmann regime).

{\vskip 0.1cm} 

Approaching the classical diluted limit yields larger and larger halo radii, galaxy masses
and velocity dispersions. On the contrary, in the quantum degenerate limit
we get solutions of the Thomas--Fermi equations corresponding to the {\bf minimal} halo radii, galaxy masses
and velocity dispersions. 

\medskip

The surface density 
\be\label{densuI}
\Sigma_0 \equiv  r_h  \; \rho_0  \simeq 120 \; M_\odot /{\rm pc}^2 \quad 
{\rm up ~ to } \; 10\% - 20\% \; ,
\ee
has the remarkable property of being nearly {\bf constant} and independent of 
luminosity in different galactic systems (spirals, dwarf irregular and 
spheroidals, elliptics) spanning over $14$ magnitudes in luminosity and over different 
Hubble types \citep{dona,span}. It is therefore a useful characteristic scale
to express galaxy magnitudes.

\bigskip

Our theoretical results follow by solving the self-consistent and nonlinear Poisson equation 
eq.(\ref{poisI}) which is {\bf solely} derived from the purely {\bf gravitational} interaction
of the WDM particles and their {\bf fermionic} nature.

\medskip  

The main galaxy magnitudes as the halo radius $ r_h $, mass $ M_h $, velocity dispersion and phase space 
density are analytically obtained and expressed in terms of the surface density, which is particularly appropriated 
to confront to observations over the whole range of galaxies. 

In this paper we {\bf derive} and analyze the general equation of state of galaxies which
clearly exhibits two regimes: 
(i) Large diluted galaxies for $ M_h \gtrsim 3.7 \; 10^6 \; M_\odot, \; \nu_0 \equiv [\mu(0)/T_0]
< -5 , \; T_0 > 0.017 $ K , described by 
the classical WDM Boltzmann regime and (ii) Compact dwarf galaxies for $ 1.6 \; 10^6 \gtrsim M_h \geq M_{h,min} 
\simeq 3.10 \; 10^4 \; \left(2 \, {\rm keV}/m\right)^{\! \! \frac{16}5} \; M_\odot, 
\; \nu_0 > -4 , \; T_0 < 0.011 $ K  described by the quantum fermionic regime close to the degenerate state. 

In particular, the $ T_0 = 0 $ degenerate or extreme quantum limit yields the most compact and smallest 
galaxy: with minimal mass $ M_{h,min} $ and minimal radius, and maximal phase space density.  

\medskip 

\begin{figure}
\begin{turn}{-90}
\psfrag{"RLMLrhmu120.dat"}{observed values}
\psfrag{"Mr0.dat"}{theory curve}
\includegraphics[height=12.cm,width=10.cm]{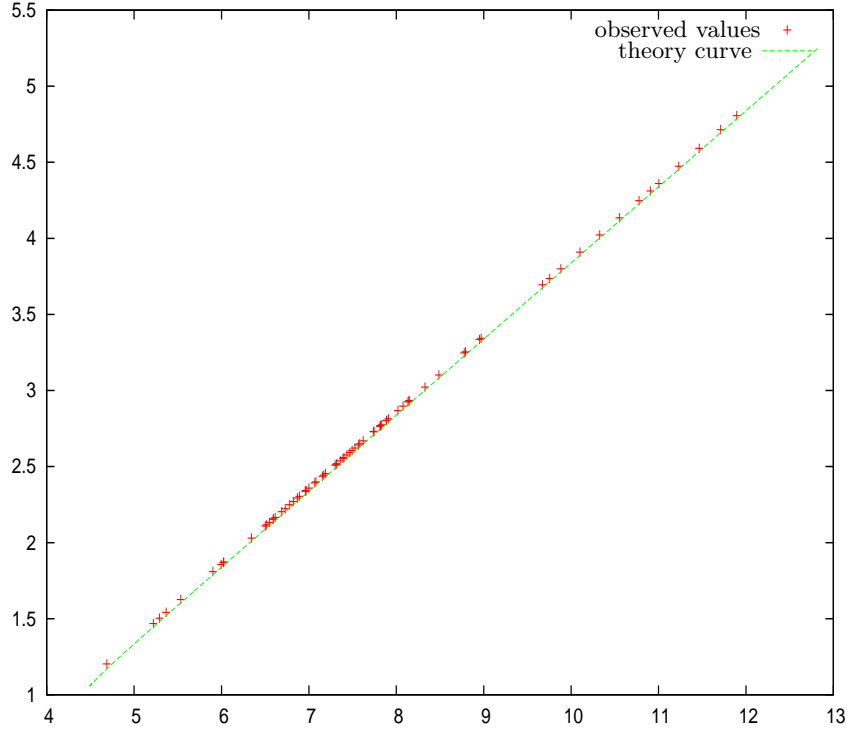}
\end{turn}
\caption{The ordinary logarithm of the halo radius $ {\hat r}_h = r_h \; \left(\frac{\Sigma_0 \;  
{\rm pc}^2}{120 \; M_\odot}\right)^{\! \! \frac15} $ vs. the ordinary logarithm of 
the halo mass $ {\hat M}_h = M_h \; \left(\frac{120 \; M_\odot}{\Sigma_0 \;  
{\rm pc}^2}\right)^{\! \! \frac35} $. We see that $ r_h $ follows with precision 
the square-root of $ M_h $ law as in the diluted regime eq.(\ref{dilu}) of the Thomas-Fermi equations. 
In addition, the galaxy data confirm the proportionality factor in this scaling relation.
The observational galaxy data for $ M_h $ and $ r_h $ are taken from Table 1 in \cite{astro} based on refs.
\cite{mcc}, \cite{sal07}, \cite{gil}, \cite{wp}, \cite{data}. The data are very well reproduced by the theoretical
Thomas-Fermi curve. The errors of the data can be estimated to be about 10-20 \%.}
\label{mhrh}
\end{figure}

We find that all magnitudes in the diluted regime exhibit square root of $ M_h $ {\bf scaling} laws and are 
{\bf universal} functions of $ r/r_h $ normalized to their values at the origin or at $ r_h $. 
Conversely, the halo mass $ M_h $ scales as the square of the halo radius $ r_h $ as
$$
 M_h = 1.75572 \; \Sigma_0 \; r_h^2 \quad .
$$
Moreover,
the proportionality factor in this scaling relation is confirmed by the galaxy data (see fig. \ref{mhrh}).

\medskip 

We find that the universal theoretical density profile obtained from the Thomas--Fermi 
equation (\ref{poisI}) in the diluted regime
($ M_h \gtrsim 10^6 \; M_\odot $) is accurately reproduced by the simple formula (see fig. \ref{ufa})
$$
\frac{\rho(r)}{\rho(0)} = \frac1{\left[1+\left(4^\frac1{\alpha}-1 \right) \; 
\left(\frac{r}{r_h}\right)^2\right]^\alpha}  
\quad , \quad \alpha = 1.5913\; .
$$
The fit being precise for $ r < 2 \; r_h $.

\medskip 

The theoretical rotation curves and density profiles obtained from the Thomas-Fermi equations 
remarkably agree with observations for $ r \lesssim r_h $, for all galaxies in the 
diluted regime \cite{urc}. This indicates that WDM is thermalized in the internal regions $ r \lesssim r_h $ of galaxies.

\medskip 

We find the WDM galaxy equation of state, that is, the functional relation between the 
pressure $ P $ and the density $ \rho $ in a parametric way as
\be\label{Ieqeg}
\rho = \frac{m^\frac52}{3 \, \pi^2 \; \hbar^3} \; \left(2 \; T_0\right)^\frac32 \; I_2(\nu)
\quad , \quad
P = \frac{m^\frac32}{15 \, \pi^2 \; \hbar^3} \; \left(2 \; T_0\right)^\frac52 \; I_4(\nu) \; .
\ee
These equations express parametrically, through the parameter $ \nu $,
the pressure $ P $ as a function of the density $ \rho $ and therefore provide the equation of state. 
$ I_2(\nu) $ and $ I_4(\nu) $ are integrals (2nd and 4th momenta) of the distribution function.
At thermal equilibrium they are given by eq.(\ref{dfI}). For the main galaxy physical magnitudes, 
the Fermi--Dirac distribution gives similar results than the 
out of equilibrium distribution functions \citep{astro}. We plot in figs. \ref{prho2} and \ref{prho},
$ P $ as a function of $ \rho $ for different values of the effective temperature $ T_0 $.

{\vskip 0.1cm} 

Interestingly enough, we provide a simple formula representing the exact equation of state (\ref{Ieqeg}) 
obtained by solving the Thomas-Fermi equation (\ref{poisI})
\be\label{Pfitrho}
P = \frac{m^\frac32\; \left(2 \; T_0\right)^\frac52 }{15 \, \pi^2 \; \hbar^3} \;\left(1 + 
\frac32 \; e^{-\beta_1 \; {\tilde \rho} } \right) \; \displaystyle 
{\tilde \rho}^{\displaystyle\frac13  \displaystyle \left(5 -
2 \;  e^{-\beta_2 \; {\tilde \rho} }\right)} \; ,
\ee
where
\be\label{Idefrhot}
{\tilde \rho} \equiv  \frac{3 \, \pi^2 \; \hbar^3}{m^\frac52 \; \left(2 \; T_0\right)^\frac32} \; \rho =I_2(\nu) \; ,
\ee
and the best fit to the exact values of $ P $ as a function $ {\tilde \rho} $ is
obtained for the values of the parameters:
\be
\beta_1 = 0.047098 \quad , \quad \beta_2 = 0.064492 \; .
\ee
The fitting formula eq.(\ref{Pfitrho}) exactly fulfils the diluted and degenerate limiting behaviours:
$$
P =  \frac{T_0}{m} \; \rho \quad 
{\rm WDM ~ diluted ~ galaxies} \quad , \quad
P = \frac{\hbar^2}5 \; \left(\frac{3 \, \pi^2}{m^4}\right)^{\! \frac23} \; \rho^\frac53 
\quad {\rm WDM ~ degenerate ~ quantum ~ limit} \; .
$$
We plot in fig. \ref{eqesfit} the exact equation of state obtained by solving
the Thomas-Fermi equation and the empirical equation of state eq.(\ref{Pfitrho}).

\medskip 

We find that the presence of universal profiles in galaxies reflect the perfect gas behaviour of the the WDM galaxy 
equation of state in the diluted regime which is identical to the self-gravitating Boltzman WDM gas. 

\medskip 

These theoretical results contrasted to robust and independent sets of galaxy data 
remarkably reproduce the observations. 

\medskip 

For the small galaxies, $ 10^6 \; M_\odot \gtrsim M_h \geq M_{h,min} $ corresponding to effective
temperatures $ T_0 \lesssim 0.017 $ K, the equation of state is steeper, dependent on the galaxy mass
and the profiles are not anymore universal. These non-universal properties in small galaxies account
to the quantum physics of the self-gravitating WDM fermions in the compact regime with high density
close to, but not at, the degenerate state. 

It would be extremely interesting to dispose of observations which could check these quantum effects in dwarf galaxies.

\medskip

In summary, the results of this paper show the power and cleanliness of the Thomas-Fermi theory and WDM
to properly describe the galaxy structures and the galaxy physical states.

\medskip

This paper is organized as follows. In Section 2 we present the Thomas-Fermi approach
to galaxy structure, we express the main galaxy magnitudes in terms of the solution
of the Thomas-Fermi equation and the value of the surface density $ \Sigma_0 $. We analyze the diluted 
classical galaxy magnitudes, derive their scaling laws and find the universal density and velocity profiles
and their agreement with observations. 

In Section 3 we derive the equation of state of galaxies and analyze their main regimes: classical regime
which is the perfect inhomogenous equation of state, identical to the WDM selfgravitating gas equation of 
state, and the quantum regime, which exhibits a steeper equation of state, non universal, galaxy mass dependent and describes 
the quantum fermionic compact states (dwarf galaxies), close to the degenerate limit. Finally, the invariance 
and dependence on the WDM particle mass $m$ in the classical and quantum regimes is discussed.

\section{Galaxy structure in the WDM Thomas-Fermi approach}\label{fortoy}

We consider DM dominated galaxies in their late stages of structure formation when 
they are relaxing to a stationary situation, at least 
not too far from the galaxy center.

{\vskip 0.1cm} 

This is a realistic situation since the free-fall (Jeans) time $ t_{ff} $ for galaxies
is much shorter than the age of galaxies:
$$
t_{ff} = \frac1{\sqrt{G \; \rho_0}} = 1.49 \; 10^7 \; 
\sqrt{\frac{M_\odot}{\rho_0 \; {\rm pc}^3}} \; {\rm yr} \; .
$$
The observed central densities of galaxies yield free-fall times in the range
from 15 million years for ultracompact galaxies till 330 million years for
large diluted spiral galaxies. These free-fall (or collapse) times are small compared
with the age of galaxies running in billions of years.

{\vskip 0.1cm} 

Hence, we can consider the DM described by a time independent and non--relativistic 
energy distribution function $ f(E) $, where $ E = p^2/(2m) - \mu $
is the single--particle energy, $ m $ is the mass of the DM particle and
$ \mu $ is the chemical potential \citep{newas,astro}
related to the gravitational potential $ \phi(\br) $ by
\be \label{potq}
  \mu(\br) =  \mu_0 - m \, \phi(\br) \; ,
\ee
where $ \mu_0 $ is a constant.

{\vskip 0.1cm} 

In the Thomas--Fermi approach, $ \rho(\br) $ is expressed as a function of $ \mu(\br) $ through the
standard integral of the DM phase--space distribution function over the momentum
\be \label{den}
  \rho(\br) = \frac{g \, m}{2 \, \pi^2 \, \hbar^3} \int_0^{\infty} dp\;p^2 
  \; f\left[\displaystyle \frac{p^2}{2m}-\mu(\br)\right] \; , 
\ee
where $ g $ is the number of internal degrees of freedom of the DM particle,
with $ g = 1 $ for Majorana fermions and $ g = 2 $ for Dirac fermions. 

\medskip

We will consider spherical symmetric configurations. Then, 
the Poisson equation for $ \phi(r) $ takes the self-consistent form
\be \label{pois}
  \frac{d^2 \mu}{dr^2} + \frac2{r} \; \frac{d \mu}{dr} = - 4\pi \, G \, m \, \rho(r) =
- \frac{2 \, g \; G \; m^2}{\pi \; \hbar^3} \int_0^{\infty} dp\;p^2 
  \; f\left[\displaystyle \frac{p^2}{2m}-\mu(r)\right]\; , 
\ee
where $ G $ is Newton's constant and $ \rho(r) $ is the DM mass density. 

\medskip

Eq. (\ref{pois}) provides an ordinary {\bf nonlinear}
differential equation that determines {\bf self-consistently} the chemical potential $ \mu(r) $ and
constitutes the Thomas--Fermi approach \citep{newas,astro} (see also ref. \citep{peter}).
This is a semi-classical approach
to galaxy structure in which the quantum nature of the DM particles is taken into account through
the quantum statistical distribution function $ f(E) $.

\medskip

The DM pressure and the velocity dispersion can also be expressed as 
integrals over the DM phase--space distribution function as
\bea \label{P}
&&  P(r) = \frac{g}{6 \, \pi^2 \,m\,\hbar^3} \int_0^{\infty} dp\;p^4 
  \,f\left[\displaystyle \frac{p^2}{2m}-\mu(r)\right] \;  , \\  \cr  \cr 
&& <v^2>(r) = \frac1{m^2} \frac{\int_0^{\infty} dp\;p^4 
  \,f\left[\displaystyle \frac{p^2}{2m}-\mu(r)\right]}{\int_0^{\infty} dp\;p^2 
  \; f\left[\displaystyle \frac{p^2}{2m}-\mu(r)\right]} \label{v2}\; .
\eea
Eqs.(\ref{den}), (\ref{P}) and (\ref{v2}) imply the equation of state
\be \label{eqest}
P(r) = \frac13 <v^2>(r) \; \rho(r) = \sigma^2(r) \; \rho(r) \; .
\ee
It must be stressed that the Thomas--Fermi equation (\ref{pois}) determine
$ \sigma^2(r) $ in terms of $ \rho(r) $ through eq.(\ref{v2}).
Therefore, the Thomas--Fermi equation {\bf determines
the equation of state} through eq.(\ref{eqest}). 
Contrary to the usual situation \citep{BT}, we {\bf do not assume}
the equation of state, but we {\bf derive} it from  the Thomas--Fermi equation.

\medskip

The fermionic DM mass density $ \rho $ is bounded at the origin 
due to the Pauli principle \citep{newas} which implies 
the bounded boundary condition at the origin
\be\label{ori}
  \frac{d \mu}{dr}(0) = 0 \; .
\ee
We see that $\mu(r)$ fully characterizes the DM halo structure in this
Thomas--Fermi framework. The chemical potential is monotonically decreasing in $ r $ 
since eq.(\ref{pois}) implies
\be\label{dmu}
\frac{d\mu}{dr} = -\frac{G\,m\,M(r)}{r^2} \quad,\qquad  
  M(r) = 4\pi \int_0^r dr'\, r'^2 \, \rho(r') \; .
\ee

From eq.(\ref{P}) and (\ref{v2}) we derive the hydrostatic equilibrium equation
\be\label{ehidr}
\frac{dP}{dr} + \rho(r) \; \frac{d\phi}{dr} = 0 \; .
\ee
Eliminating $ P(r) $ between eqs.(\ref{eqest}) and (\ref{ehidr}) and integrating on $ r $ gives
\be\label{bariot}
\frac{\rho(r)}{\rho(0)} = \frac{\sigma^2(0)}{\sigma^2(r)} \; 
e^{ \displaystyle -\int_0^r\frac{dr'}{\sigma^2(r')} \; \frac{d\phi}{dr'}} \; .
\ee
Inserting this expression in the Poisson's equation yields
\be\label{ecgral}
\frac{d^2 \phi}{dr^2} + \frac2{r} \; \frac{d \phi}{dr} =  4 \, \pi \; G \; \rho_0 \;
\frac{\sigma^2(0)}{\sigma^2(r)} \; e^{- \displaystyle \int_0^r\frac{dr'}{\sigma^2(r')} \; 
\frac{d\phi}{dr'}} \; .
\ee
This nonlinear equation for non constant $ \sigma^2(r) $
generalizes the corresponding equation in the self-gravitating Boltzmann gas.
For constant $ \sigma^2(r) $ eq.(\ref{bariot}) reduces to the baryotropic equation.

{\vskip 0.2cm} 

In this semi-classical framework the stationary energy distribution function $
f(E) $ must be given. We consider the Fermi--Dirac distribution 
\be\label{FD}
  f(E) = \Psi_{\rm FD}(E/T_0) = \frac1{e^{E/T_0} + 1} \; ,
\ee
where the characteristic one--particle energy scale $ T_0 $ in the DM halo
plays the role of an effective temperature. The value of $ T_0 $ depends on the galaxy mass.
In neutron stars, where the neutron mass is about six orders of magnitude larger
than the WDM particle mass, the temperature can be approximated by zero.
In galaxies, $ T_0 \sim m \; <v^2> $ turns to be non-zero but small in the range: 
$ 10^{-3} \; {\rm K} \lesssim T_0  \lesssim 10 $ K for halo galaxy masses in 
the range $ 10^5 - 10^{12} \; M_\odot $ which reproduce the observed velocity
dispersions for $ m \simeq 2 $ keV. The smaller values of $ T_0 $ correspond to compact
dwarfs and the larger values of $ T_0 $ are for large and diluted galaxies.

{\vskip 0.1cm} 

Notice that for the relevant galaxy physical magnitudes,
the Fermi--Dirac distribution give similar results than the 
out of equilibrium distribution functions \citep{astro}.

{\vskip 0.1cm} 

The choice of $ \Psi_{\rm FD} $ is justified in the inner regions,
where relaxation to thermal equilibrium is possible. Far from the  origin
however, the Fermi--Dirac distribution as its classical counterpart, the isothermal sphere,
produces a mass density tail $ 1/r^2 $ that overestimates the observed tails of the 
galaxy mass densities. Indeed, the
classical regime $ \mu/T_0 \to -\infty $ is attained for large distances $ r $
since eq.(\ref{dmu}) indicates that $ \mu(r) $ is always monotonically decreasing with $ r $.

{\vskip 0.1cm} 

More precisely, large positive values of the chemical potential at the origin 
correspond to the degenerate fermions limit which is the extreme quantum case and oppositely, 
large negative values of the chemical potential at the origin gives the diluted case which 
is the classical regime. The quantum degenerate regime describes dwarf and compact galaxies while
the classical and diluted regime describes large and diluted galaxies.
In the classical regime, the Thomas-Fermi equation
(\ref{pois})-(\ref{ori}) become the equations for a self-gravitating Boltzmann gas.

\medskip

It is useful to introduce dimensionless variables $ \xi , \; \nu(\xi) $ 
\be\label{varsd}
 r = l_0 \; \xi \quad , \quad \mu(r) =  T_0 \;  \nu(\xi) \quad , \quad f(E) = \Psi(E/T_0) \; , 
\ee
where $ l_0 $ is the characteristic length that emerges from the dynamical equation (\ref{pois}):
\be\label{varsd2}
l_0 \equiv  \frac{\hbar}{\sqrt{8\,G}} \; \left(\frac2{g}\right)^{\! \! \frac13} \;
\left[\frac{9 \, \pi \; I_2(\nu_0)}{m^8\,\rho_0}\right]^{\! \! \frac16} 
  = R_0 \; \left(\frac{2 \, {\rm keV}}{m}\right)^{\! \! \frac43}  \; \left(\frac2{g}\right)^{\! \! \frac13} \;
  \left[\frac{I_2(\nu_0)}{\rho_0} \; \frac{M_\odot}{{\rm pc}^3}\right]^{\! \! \frac16} 
  \;,\qquad R_0 = 7.425 \; \rm pc  \; ,
\ee
and
\be\label{dfI}
I_n(\nu) \equiv (n+1) \; \int_0^{\infty} y^n \; dy \; \Psi_{FD}(y^2 -\nu) \quad  , \quad
n = 1, 2 , \ldots\; , \quad  , \quad \nu_0 \equiv \nu(0) \quad  , \quad \rho_0 = \rho(0) \; ,
\ee
where we use the integration variable $ y \equiv p / \sqrt{2 \, m \;  T_0} $.
For definiteness, we will take $ g=2 $, Dirac fermions in the sequel. 
One can easily translate from Dirac to Majorana fermions changing the WDM fermion mass as:
$$
m \Rightarrow \frac{m}{2^\frac14} = 0.8409 \; m \; .
$$

\medskip

Then, in dimensionless variables, the self-consistent Thomas-Fermi equation 
(\ref{pois}) for the chemical potential $ \nu(\xi) $ takes the form
\be\label{nu}
\frac{d^2 \nu}{d\xi^2} + \frac2{\xi} \; \frac{d \nu}{d\xi} = - I_2(\nu)
\quad ,  \quad \nu'(0) = 0 \quad .
\ee

We find the main physical galaxy magnitudes, such as the
mass density $ \rho(r) $, the velocity dispersion $ \sigma^2(r) = v^2(r)/3 $ and the pressure 
$ P(r) $, which are all $r$-dependent as: 
\bea\label{gorda}
&& \rho(r) = \frac{m^\frac52}{3 \, \pi^2 \; \hbar^3} \; \left(2 \; T_0\right)^\frac32 \; I_2(\nu(\xi)) =
\rho_0 \; \frac{I_2(\nu(\xi))}{I_2(\nu_0)} 
\quad , \quad  \rho_0 = \frac{m^\frac52}{3 \, \pi^2 \; \hbar^3} \; \left(2 \; T_0\right)^\frac32 \; I_2(\nu_0) \; ,
 \\ \cr \cr 
&& P(r) =  \frac{m^\frac32}{15 \, \pi^2 \; \hbar^3} \; \left(2 \; T_0\right)^\frac52 \; I_4(\nu(\xi))
= \frac1{5} \; \left(9 \; \pi^4\right)^{\!\frac13} \; \left(\frac{\hbar^6}{m^8}\right)^{\! \frac13} \;
\left[\frac{\rho_0}{I_2(\nu_0)}\right]^{5/3} \; I_4(\nu(\xi)) \;, \label{pres1}\\ \cr \cr 
&& \sigma^2(r) = \frac{P(r)}{\rho(r)} =
\frac{2 \; T_0}{5 \; m} \; \frac{I_4(\nu(\xi))}{I_2(\nu(\xi))}  \quad .\label{sig2}
\eea
As a consequence, from eqs.(\ref{dmu}), (\ref{varsd}), (\ref{varsd2}), (\ref{gorda}) and (\ref{sig2})
the total mass $ M(r) $ enclosed in a sphere of radius $ r $ and
the phase space density $ Q(r) $ turn to be
\bea\label{cero} 
&& M(r) = 4 \, \pi \; \frac{\rho_0\; l_0^3}{I_2(\nu_0)}\,\int_0^{\xi}
    dx\, x^2 \,I_2(\nu(x)) = 4 \, \pi \; \frac{\rho_0 \; l_0^3}{I_2(\nu_0)} \;
    \xi^2 \; |\nu'(\xi)| = \\ \cr \cr
&&=  M_0 \; \xi^2 \; |\nu'(\xi)|
    \; \left(\frac{{\rm keV}}{m}\right)^{\! \! 4} \; 
\sqrt{\frac{\rho_0}{I_2(\nu_0)}\; \frac{{\rm pc}^3}{M_\odot}} \; , 
\quad  M_0 = 4 \; \pi \; M_\odot \; \left(\frac{R_0}{\rm pc}\right)^{\! \! 3} 
    = 0.8230 \; 10^5 \; M_\odot \; ,\label{emeder} \\ \cr 
&& Q(r) \equiv \frac{\rho(r)}{\sigma^3(r)} = 3 \; \sqrt3 \; \frac{\rho(r)}{<v^2>^\frac32(r)} 
= \frac{\sqrt{125}}{3 \; \pi^2} \; \; \frac{m^4}{\hbar^3}  \;
\frac{I_2^{\frac{5}{2}}(\nu(\xi))}{I_4^{\frac32}(\nu(\xi))} \label{Qcero} \; .
\eea
In these expressions, we have systematically eliminated the energy scale $ T_0 $ 
in terms of the central density $ \rho_0 $ through eq.(\ref{gorda}). 
Notice that $ Q(r) $ turns to be independent of $ T_0 $ and therefore of $ \rho_0 $.

\medskip

We define the core size $ r_h $ of the halo by analogy with the Burkert density profile as
\be\label{onequarter}
  \frac{\rho(r_h)}{\rho_0} = \frac14 \quad , \quad  r_h = l_0 \; \xi_h \; .
\ee

\medskip

It must be noticed that the surface density 
 \be\label{densu}
\Sigma_0 \equiv  r_h  \; \rho_0  \; ,
\ee
is found nearly {\bf constant} and independent  of 
luminosity in  different galactic systems (spirals, dwarf irregular and 
spheroidals, elliptics) 
spanning over $14$ magnitudes in luminosity and  over different 
Hubble types. More precisely, all galaxies seem to have the same value 
for $ \Sigma_0 $, namely $ \Sigma_0 \simeq 120 \; M_\odot /{\rm pc}^2 $
up to $ 10\% - 20\% $ \citep{dona,span,kor}.   
It is remarkable that at the same time 
other important structural quantities as $ r_h , \; \rho_0 $, 
the baryon-fraction and the galaxy mass vary orders of magnitude 
from one galaxy to another.

{\vskip 0.1cm} 

The constancy of $ \Sigma_0 $ seems unlikely to be a mere coincidence and probably
reflects a physical scaling relation between the mass and halo size of galaxies.
It must be stressed that $ \Sigma_0 $ is the only dimensionful quantity
which is constant among the different galaxies.

\medskip

It is then useful to take here the dimensionful quantity $ \Sigma_0 $ as physical scale 
to express the galaxy magnitudes in the Thomas-Fermi approach. 
That is, we replace the central density $ \rho_0 $
in the above galaxy magnitudes eqs.(\ref{varsd2})-(\ref{emeder}) in terms of 
$ \Sigma_0 $ eq.(\ref{densu}) with the following results
\bea\label{E0}
&& l_0 = \left(\frac{9 \; \pi}{2^9}\right)^{\! \! \frac15} \; 
\left(\frac{\hbar^6}{G^3 \; m^8}\right)^{\! \! \frac15} \; 
\; \left[\frac{\xi_h \; I_2(\nu_0)}{\Sigma_0}\right]^{\! \! \frac15} =
4.2557 \; \left[\xi_h \; I_2(\nu_0)\right]^{\! \frac15} \;
\left(\frac{2 \, {\rm keV}}{m}\right)^{\! \! \frac85} \; 
\left(\frac{120 \; M_\odot}{\Sigma_0 \;  {\rm pc}^2}\right)^{\! \! \frac15}  \;
{\rm pc}\; , \\ \cr\cr
&& T_0 = \left(18 \; \pi^6 \; \frac{\hbar^6 \; G^2}{m^3}\right)^{\! \frac15}
\; \left[\frac{\Sigma_0}{\xi_h \; I_2(\nu_0)}\right]^{\! \! \frac45} = \frac{7.12757 \; 10^{-3}}{
\left[\xi_h \; I_2(\nu_0)\right]^{\frac45}} \;  \left(\frac{2 \, {\rm keV}}{m}\right)^{\! \! \frac35} \; 
\left(\frac{\Sigma_0 \;  {\rm pc}^2}{120 \; M_\odot}\right)^{\! \! \frac45}  \; {\rm K} \; , \label{T0}
\eea
and
\bea\label{rMr}
&& r = 4.2557 \;  \xi \; \left[ \xi_h \; I_2(\nu_0) \right]^{\frac15} \;
 \left(\frac{2 \, {\rm keV}}{m}\right)^{\! \! \frac85}  \; 
\left(\frac{120 \; M_\odot}{\Sigma_0 \;  {\rm pc}^2}\right)^{\! \! \frac15} \; {\rm pc} \; , \\ \cr\cr
&& \rho(r) = \left(\frac{2^9 \; G^3 \; m^8}{9 \; \pi \; \hbar^6}\right)^{\! \! \frac15} \; 
\left[\frac{\Sigma_0}{\xi_h \; I_2(\nu_0)}\right]^{\! \frac65} \; I_2(\nu(\xi))=
 18.1967 \; \frac{I_2(\nu(\xi))}{\left[\xi_h \; I_2(\nu_0)\right]^{\! \frac65}}
\; \left(\frac{m}{2 \, {\rm keV}}\right)^{\! \! \frac85} \; 
\left(\frac{\Sigma_0 \;  {\rm pc}^2}{120 \; M_\odot}\right)^{\! \! \frac65} \;
\frac{M_\odot}{{\rm pc}^3} \; , \label{rhor}\\ \cr\cr
&& M(r) = 4 \, \pi \;  
\left(\frac{9 \; \pi \; \hbar^6}{2^9 \; G^3 \; m^8}\right)^{\! \! \frac25} \; 
\left[\frac{\Sigma_0}{\xi_h \; I_2(\nu_0)}\right]^{\! \! \frac35}
\; \xi^2 \; |\nu'(\xi)| 
=\frac{27312 \; \xi^2}{\left[ \xi_h \; I_2(\nu_0) \right]^{\frac35}}
\; |\nu'(\xi)| \; \left(\frac{2 \, {\rm keV}}{m}\right)^{\! \! \frac{16}5}
\left(\frac{\Sigma_0 \; {\rm pc}^2}{120 \; M_\odot}\right)^{\! \! \frac35}  
M_\odot \; , \label{emer} \\ \cr\cr
&& \sigma^2(r) = \frac{11.0402}{\left[\xi_h \; I_2(\nu_0)\right]^{\! \frac45}} \; 
\frac{I_4(\nu(\xi))}{I_2(\nu(\xi))} \; 
\left(\frac{2 \, {\rm keV}}{m}\right)^{\! \! \frac85} \;
\left(\frac{\Sigma_0 \;  {\rm pc}^2}{120 \; M_\odot}\right)^{\! \! \frac45}  \;
\left(\frac{\rm km}{\rm s}\right)^2 \; \; , \\ \cr\cr
&& P(r) = \frac{8 \, \pi}5 \; G \; \left[\frac{\Sigma_0}{\xi_h \; I_2(\nu_0)}\right]^2
\;  I_4(\nu(\xi))
= \frac{200.895}{\left[\xi_h \; I_2(\nu_0)\right]^2} \; I_4(\nu(\xi)) \; 
\left(\frac{\Sigma_0 \;  {\rm pc}^2}{120 \; M_\odot}\right)^2 \; \frac{M_\odot}{{\rm pc}^3} \;
 \left(\frac{\rm km}{\rm s}\right)^2 \; \; .\label{presion}
\eea
For a fixed value of the surface density $ \Sigma_0 $, the 
solutions of the Thomas-Fermi eqs.(\ref{nu}) are parametrized by a single
parameter: the dimensionless chemical potential at the origin $ \nu_0 $. The value of
$ \nu_0 $ at fixed $ \Sigma_0 $ can be determined by the value of the halo galaxy 
mass $ M(r_h) $ obtained from eq.(\ref{emer}) at $ r = r_h $. 
\be\label{Mrh}
M_h \equiv  M(r_h) = \frac{27312 \; \xi_h^{\frac75}}{\left[I_2(\nu_0) \right]^{\frac35}}
\; |\nu'(\xi_h)| \; \left(\frac{2 \, {\rm keV}}{m}\right)^{\! \! \frac{16}5}
\left(\frac{\Sigma_0 \; {\rm pc}^2}{120 \; M_\odot}\right)^{\! \! \frac35}  M_\odot \; .
\ee
Also, at fixed surface density $ \Sigma_0 $, the effective temperature $ T_0 $ is only a function of $ \nu_0 $.

{\vskip 0.2cm}

It is useful to introduce the rescaled dimensionless variables
\bea\label{rsom}
&& {\hat r}_h \equiv r_h \; \left(\frac{m}{2 \, {\rm keV}}\right)^{\! \! \frac85} \;  
\left(\frac{\Sigma_0 \; {\rm pc}^2}{120 \; M_\odot}\right)^{\! \! \frac15} 
\quad , \quad 
{\hat M}_h \equiv M_h  \; \left(\frac{m}{2 \, {\rm keV}}\right)^{\! \! \frac{16}5} \; 
\left(\frac{120 \; M_\odot}{\Sigma_0 \; {\rm pc}^2}\right)^{\! \! \frac35}  \quad , \quad {\hat T}_0 \equiv T_0 \; 
\frac{2 \, {\rm keV}}{m} \; \left(\frac{120 \; M_\odot}{\Sigma_0 \; {\rm pc}^2}\right)^{\! \! \frac45} \; 
\cr \cr 
&& {\hat \nu}_0 \equiv  \nu_0 + 4 \; \ln\left(\frac{m}{2 \, {\rm keV}}\right) \quad , \quad 
{\hat\sigma}^2(r) \equiv \sigma^2(r) \; \left(\frac{m}{2 \, {\rm keV}}\right)^{\! \! \frac85} \;
\left(\frac{120 \; M_\odot}{\Sigma_0 \;  {\rm pc}^2}\right)^{\! \! \frac45} 
\quad , 
\eea
We display in Table \ref{MhT0nu0} the corresponding values of the halo mass $ {\hat M}_h $, the effective temperature
$ {\hat T_0} $ and the chemical potential at the origin $ \nu_0 $
in the whole galaxy mass range, from large diluted galaxies till small ultracompact galaxies.

\medskip

The circular velocity $ v_c(r) $ is defined through the virial theorem as
\be\label{vci}
 v_c(r) \equiv \sqrt{\frac{G \; M(r)}{r}} \; ,
\ee
and it is directly related by eq.(\ref{dmu}) 
to the derivative of the chemical potential as
$$
v_c(r) = \sqrt{- \frac{r}{m} \; \frac{d\mu}{dr}} = \sqrt{-\frac{T_0}{m} \; \frac{d\nu}{d\ln \xi}} \; .
$$
Expressing $ T_0 $ in terms of the surface density  $ \Sigma_0 $ using
eq.(\ref{T0}) we have for the circular velocity the explicit expression
\be\label{vtf}
v_c(r) = 5.2537 \; \frac{\sqrt{-\xi \; \nu'(\xi)}}{\left[ \xi_h \; I_2(\nu_0) \right]^{\frac25}}
\; \left(\frac{2 \, {\rm keV}}{m}\right)^{\! \! \frac45} \;
\left(\frac{\Sigma_0 \;  {\rm pc}^2}{120 \; M_\odot}\right)^{\! \! \frac25}  \;
\frac{\rm km}{\rm s} \; .
\ee
and the rescaled circular velocity,
$$
{\hat v}_c^2(r) \equiv v_c^2(r)
\; \left(\frac{m}{2 \, {\rm keV}}\right)^{\! \! \frac85} \;
\left(\frac{120 \; M_\odot}{\Sigma_0 \;  {\rm pc}^2}\right)^{\! \! \frac45} \; . \nonumber
$$
\medskip

Two important combinations of galaxy magnitudes are $ r \; \rho(r') $ and $ M(r)/[4 \, \pi \; r^2] $.
From eqs. (\ref{varsd}), (\ref{cero}), (\ref{E0}) and (\ref{rhor}) we obtain
\be\label{denssr}
r \; \rho(r') = \Sigma_0 \; \frac{\xi \; I_2(\nu(\xi'))}{\xi_h \; I_2(\nu_0)}  \quad ,
\quad \frac{M(r)}{4 \, \pi \; r^2} = \Sigma_0 \; \frac{|\nu'(\xi)|}{\xi_h \; I_2(\nu_0)} \; .
\ee
In particular, it follows that $ r_h \; \rho(0) = \Sigma_0 $ reproducing the surface density as it must be.
At a generic point $ r $ eqs.(\ref{denssr}) provide expressions for a space dependent
surface density. They are both proportional to $ \Sigma_0 $ and differ from each other by
factors of order one. Notice that $ \hbar, \; G $ and $ m $ canceled out in 
these space dependent surface densities eqs.(\ref{denssr}).

\begin{figure*}
\begin{turn}{-90}
\psfrag{"MT.dat"}{Exact Thomas-Fermi}
\psfrag{"MTlin.dat"}{diluted regime}
{\label{MT}\includegraphics[height=12.cm,width=7.cm]{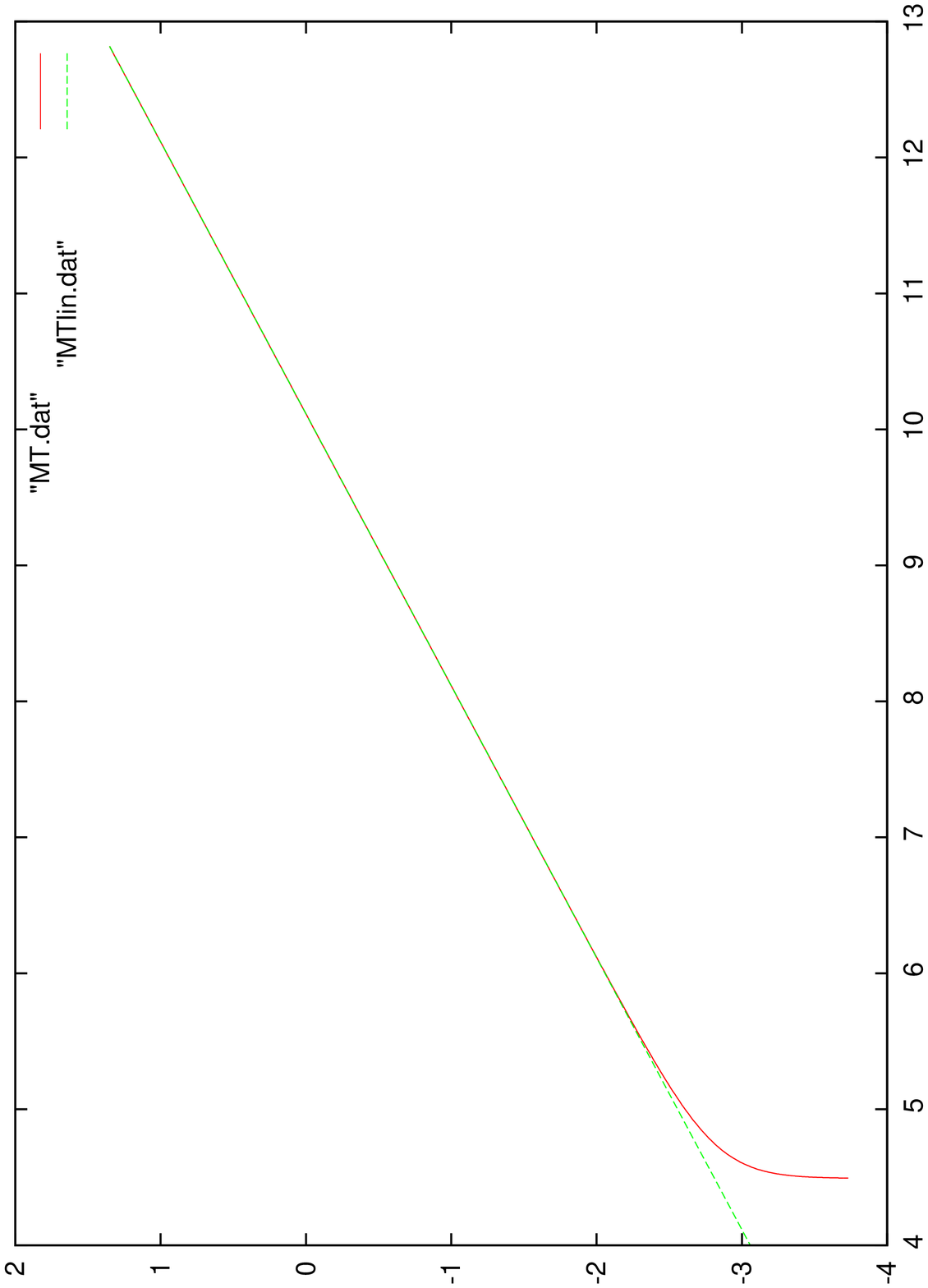}}
\psfrag{"Mnu0.dat"}{Exact Thomas-Fermi}
\psfrag{"Mnu0lin.dat"}{diluted regime}
{\label{Mnu0}\includegraphics[height=12.cm,width=7.cm]{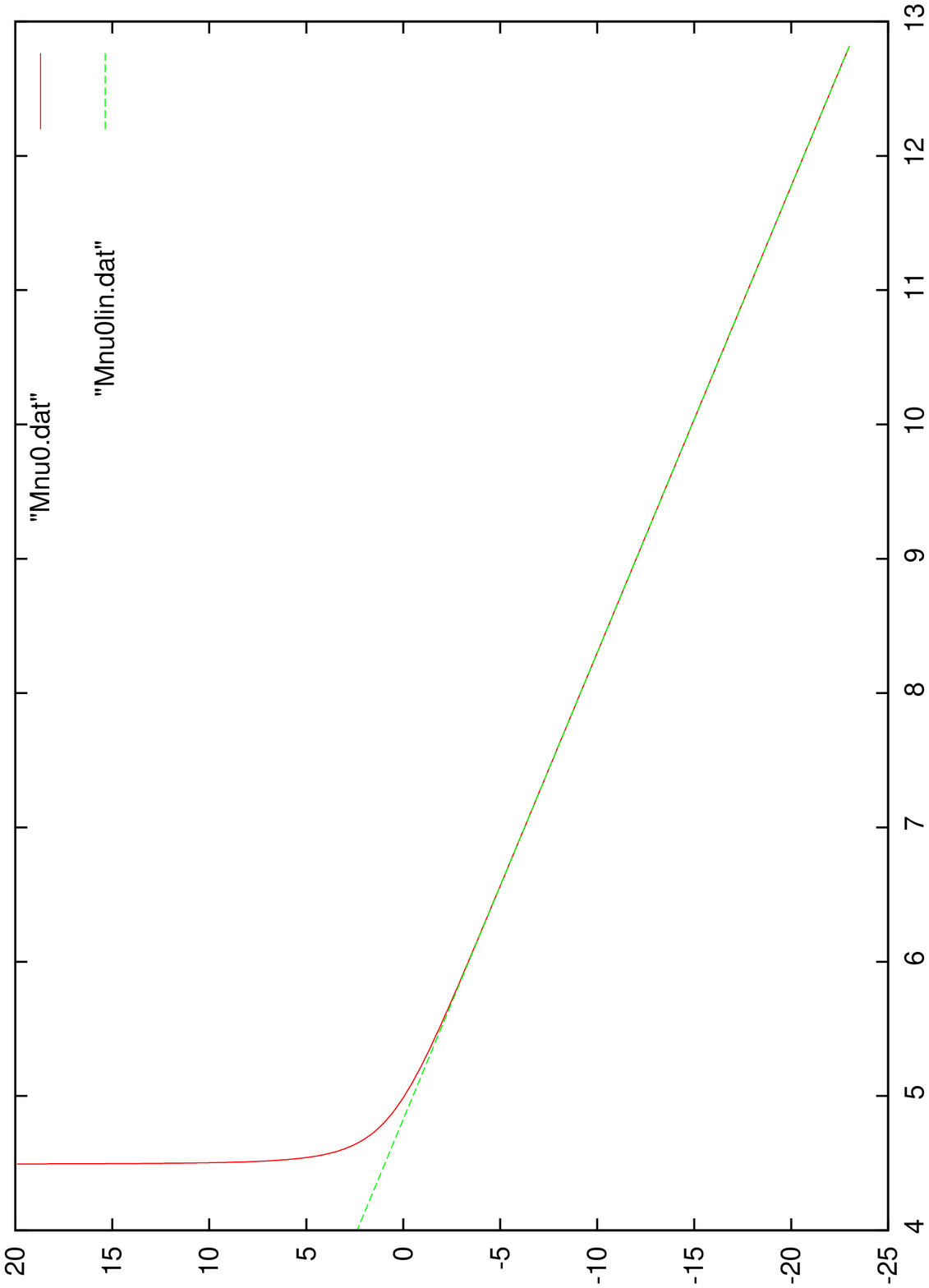}}
\end{turn}
\caption{Upper Panel: The ordinary logarithm of the effective temperature
$ {\hat T}_0 $ vs. the ordinary logarithm of the halo mass $ {\hat M}_h $.
We see that $ {\hat T}_0  $ grows with $ {\hat M}_h $ following with precision 
the square-root of $ {\hat M}_h $ law as in the diluted regime eq.(\ref{dilu}) of the Thomas-Fermi equations,
except for $ {\hat M}_h < 3 \; 10^5 \; M_\odot, \; \nu_0 > -1.8, \; {\hat T_0} < 0.005 $ K 
which is near the quantum degenerate regime
and corresponds to compact dwarf galaxies. The deviation from the scaling diluted regime is due to the quantum 
fermionic effects which become important for dwarf compact galaxies. Lower Panel.
The dimensionless chemical potential at the origin $ {\hat \nu}_0 $
vs. the ordinary logarithm of $ {\hat M}_h $. We see that $ \nu_0  $ follows with precision 
the $ (5/4) \log{\hat M}_h $ law as in the diluted regime eq.(\ref{dilu}) of the Thomas-Fermi equations. 
except near the degenerate regime for $ {\hat M}_h < 3 \; 10^5 \; M_\odot, \; \nu_0 > -1.8, \; {\hat T_0} < 0.005 $ K 
corresponding to compact dwarf galaxies.}
\label{Tb}
\end{figure*}

\begin{table}
\begin{tabular}{|c|c|c|} \hline  
 & &   \\
 $ \displaystyle {\hat M}_h $ &  $ \displaystyle {\hat T_0} $  & $ \quad \nu_0 = \displaystyle \frac{\mu(0)}{ T_0} \quad $
 \\ & & \\ \hline  
$ 6.56 \; 10^{12} \; M_\odot $ & 22.4 \; K  & -23 
\\ \hline  
$ \quad 6.45 \; 10^{11} \; M_\odot \quad $ & 7.04 \; K & -20.1   \\  
\hline  
$ 6.34 \; 10^{10} \; M_\odot $ & 2.21 \; K & -17.2  \\ 
\hline  
$ 4.9 \; 10^9 \; M_\odot $ & 0.613 \; K & -14 \\
\hline  
$ 2.16 \; 10^8 \; M_\odot $ & 0.129 \; K & -10.1  \\ 
\hline  
$ 1.55 \; 10^7\; M_\odot  $ & 0.0344 \; K & -6.8 \\ 
\hline  
$ 3.67 \; 10^6\; M_\odot  $ & 0.0168 \; K & -5  \\ 
\hline  
$ 1.66 \; 10^6\; M_\odot  $ & 0.0112 \; K & -4  \\ 
\hline  
$ 1.21 \; 10^5\; M_\odot  $ & 0.00278 \; K & -0.4  \\ 
\hline  
$ 9.73 \; 10^4\; M_\odot  $ & 0.00241 \; K & 0  \\ 
\hline  
$ 6.31 \; 10^4\; M_\odot  $ & 0.00173 \; K & 1  \\ 
\hline  
$ 4.06 \; 10^4\; M_\odot  $ & 0.00101 \; K & 3  \\ 
\hline  
$ 3.48 \; 10^4\; M_\odot  $ & $ 6.82 \; 10^{-4} \; $ K  & 5  \\ 
\hline  
$ 3.19 \; 10^4\; M_\odot  $ & $ 3.63 \; 10^{-4} \; $ K & 10  \\ 
\hline  
$ 3.12 \; 10^4\; M_\odot  $ & $ 1.84 \; 10^{-4} \; $ K & 20  \\ 
\hline  
$  {\hat M}_h^{min} = 3.10 \; 10^4\; M_\odot $ & 0  & $ +\infty $ 
\\ \hline  
\end{tabular}
\caption{Corresponding values of the halo mass $ {\hat M}_h $, the effective temperature
$ {\hat T_0} $ and the chemical potential at the origin $ \nu_0 $
for WDM galaxies covering the whole range from large diluted galaxies till small ultracompact galaxies.}
\label{MhT0nu0}
\end{table}

\subsection{Galaxy properties in the diluted Boltzmann regime}

In the diluted Boltzmann regime, $ \nu_0 \lesssim -5 $, the analytic expressions
for the main galaxies magnitudes are given by:
\bea\label{dilu}
&&  M_h = 1.75572 \; \Sigma_0 \; r_h^2 \quad , \quad 
r_h =  68.894 \; \sqrt{\frac{M_h}{10^6 \; M_\odot}
\frac{120\; M_\odot}{\Sigma_0 \;  {\rm pc}^2}} \;\; \; {\rm pc}\; ,  \\ \cr \cr
&& T_0 = 8.7615 \; 10^{-3} \; \sqrt{\frac{M_h}{10^6 \; M_\odot}} \; \frac{m}{2 \, {\rm keV}}
\; \sqrt{\frac{\Sigma_0 \;  {\rm pc}^2}{120 \; M_\odot}} \; {\rm K} \quad , \\ \cr \cr
&& \rho(r) = 5.19505 \; \left(\frac{M_h}{10^4 \; M_\odot} \; \frac{\Sigma_0 \;  {\rm pc}^2}{120 \; M_\odot}
\right)^{\! \! \frac34} \; \left(\frac{m}{2 \, {\rm keV}}\right)^4 \; e^{\nu(\xi)} \; \; 
\frac{M_\odot}{{\rm pc}^3} \; ,\\ \cr \cr
&& \sigma^2(r) = 33.927 \; \sqrt{\frac{M_h}{10^6 \; M_\odot} \;
\frac{\Sigma_0 \;  {\rm pc}^2}{120 \; M_\odot} } \;
\left(\frac{\rm km}{\rm s}\right)^2 \; \; ,\label{sigmac} \\ \cr \cr
&&  v_c^2(r) = 33.9297 \; \sqrt{\frac{M_h}{10^6 \; M_\odot} \;
\frac{\Sigma_0 \;  {\rm pc}^2}{120 \; M_\odot}} \;
\left| \frac{d \nu(\xi)}{d \ln \xi}\right| 
\; \left(\frac{\rm km}{\rm s}\right)^2 \; , \; 
v_c^2(r_h) = 62.4292 \; \sqrt{\frac{M_h}{10^6 \; M_\odot} \;
\frac{\Sigma_0 \;  {\rm pc}^2}{120 \; M_\odot}} \;
\; \left(\frac{\rm km}{\rm s}\right)^2 \; \label{vch} \\ \cr \cr
&& M(r) = 7.88895 \; \left| \frac{d \nu(\xi)}{d \ln \xi}\right|  \; \frac{r}{\rm pc}
\; \sqrt{\frac{M_h}{10^6 \; M_\odot} \; \frac{\Sigma_0 \;  {\rm pc}^2}{120 \; M_\odot}} \; .
\eea
In addition, 
$ M_h $ and $ T_0 $ scale as  functions of the fugacity at the center $ z_0 = e^{\nu_0} $:
\bea\label{escanu0}
&&  M_h \equiv M(r_h) 
= \frac{67014.6}{z_0^\frac45} \; \left(\frac{2 \, {\rm keV}}{m}\right)^{\! \! \frac{16}5} \;
 \left(\frac{\Sigma_0 \;  {\rm pc}^2}{120 \; M_\odot}\right)^{\! \! \frac35}  \;  M_\odot \; ,\\ \cr \cr
&& T_0 = \frac{2.2681 \; 10^{-3}}{z_0^\frac25} \; \left(\frac{2 \, {\rm keV}}{m}\right)^{\! \! \frac35}  \; 
\left(\frac{\Sigma_0 \;  {\rm pc}^2}{120 \; M_\odot}\right)^{\! \! \frac45} \; {\rm K} \label{T0dil} \; .
\eea
Therefore, {\bf all} these galaxy magnitudes {\bf scale} as functions of each other.

{\vskip 0.2cm}

For the equation of state and the phase space density we find the expressions
\bea
&& P(r) = 5.57359 \; 10^3 \; \left(\frac{M_h}{10^6 \; M_\odot} \; \frac{\Sigma_0 \;  {\rm pc}^2}{120 \; M_\odot}
\right)^{\! \! \frac54} \; \left(\frac{m}{2 \, {\rm keV}}\right)^4 \; e^{\nu(\xi)} \; \; \frac{M_\odot}{{\rm pc}^3} \;
\left(\frac{\rm km}{\rm s}\right)^2 \; , \\ \cr \cr
&& P_0 \equiv P(0) = 59.097 \; \left(\frac{\Sigma_0 \;  {\rm pc}^2}{120 \; M_\odot}\right)^2 
\; \frac{M_\odot}{{\rm pc}^3} \; \left(\frac{\rm km}{\rm s}\right)^2 \; , \\ \cr \cr
&& Q(r) = 2.031796 \; \left(\frac{m}{2 \, {\rm keV}}\right)^4 \; e^{\nu(\xi)} \; 
 {\rm keV}^4 \quad , \quad 
Q(0) = 1.2319 \; \left(\frac{10^5 \;  M_\odot}{M_h}\right)^{\! \! \frac54} \;
\left(\frac{\Sigma_0 \;  {\rm pc}^2}{120 \; M_\odot}\right)^{\! \! \frac34} \; {\rm keV}^4 \; \label{vcird}\; .
\eea
These equations are accurate for $ M_h \gtrsim 10^6 \;  M_\odot $.
We see that they exhibit a {\bf scaling} behaviour for $ r_h, \;  T_0, \;   Q(0) , \; \sigma^2(0) $
and  $ v_c^2(r_h) $ as functions of $ M_h $. 

{\vskip 0.2cm} 

We see from eqs.(\ref{dilu}) and (\ref{sigmac}) that $ T_0 $ and $ m \; \sigma^2(0) $ only 
differ by purely numerical factors reflecting the equipartition of kinetic energy.
More precisely, it follows from eqs.(\ref{dilu}) and (\ref{sigmac}) that
\be\label{equip}
\frac{m}2 \; <v^2(0)> \; = \frac32 \; m \; \sigma^2(0) = \frac32 \; T_0 \; ,
\ee
which shows that in the diluted regime the self-gravitating WDM gas behaves as an 
{\bf inhomogeneous perfect gas} as we will discuss in the next section.

{\vskip 0.2cm}

We plot in figs. \ref{Tb} and \ref{mhrh}, the dimensionless effective temperature $ {\hat T}_0 $, 
the chemical potential at the origin $ {\hat \nu}_0 $ and the normalized halo radius 
$ {\hat r}_h $ as functions of the halo mass $ \hat M_h $ as defined by eqs.(\ref{rsom}). 
We also depict in fig. \ref{mhrh} the galaxy observations from different sets of data from refs. 
\cite{mcc}, \cite{sal07}, \cite{gil}, \cite{wp}, \cite{data}. 
All data are well reproduced by 
our theoretical Thomas-Fermi results. The errors of the data can be estimated to be about 10-20 \%.

{\vskip 0.1cm} 

The characteristic temperature $ \hat T_0 $ monotonically grows with the halo mass 
$ {\hat M}_h $ of the galaxy as shown by fig. \ref{Tb} and eq.(\ref{escanu0}) following
with good precision the square root of $ {\hat M}_h $ eq.(\ref{dilu}).

{\vskip 0.1cm} 

We see that the whole set of scaling behaviours 
of the diluted regime eqs. (\ref{dilu})-(\ref{escanu0})
are {\bf very accurate} except near the degenerate regime for halo masses $ {\hat M}_h < 3 \; 10^5 \; M_\odot $.
The deviation from the diluted scaling regime for $ {\hat M}_h < 3 \; 10^5 \; M_\odot $
accounts for the quantum fermionic effects in the dwarf compact galaxies obtained in our Thomas-Fermi
approach.

{\vskip 0.1cm} 

It must be stressed that the scaling relations eqs.(\ref{dilu})-(\ref{vcird}) 
are a consequence solely of the self-gravitating interaction of the fermionic WDM.
Galaxy data verify the exponent and the amplitude factor in these scaling as shown 
in fig. \ref{mhrh} for the square root scaling relation eq.(\ref{dilu}).

{\vskip 0.1cm} 

It is highly remarkable that our theoretical
results {\bf reproduce} the observed DM halo properties with {\bf good precision}.

\begin{figure}
\begin{turn}{-90}
\psfrag{"vsT.dat"}{$ {\hat v}^2(0) = 3 \; {\hat \sigma}^2(0) $}
\psfrag{"vcirT.dat"}{$ {\hat v}_c^2(r_h) $}
\includegraphics[height=12.cm,width=7.cm]{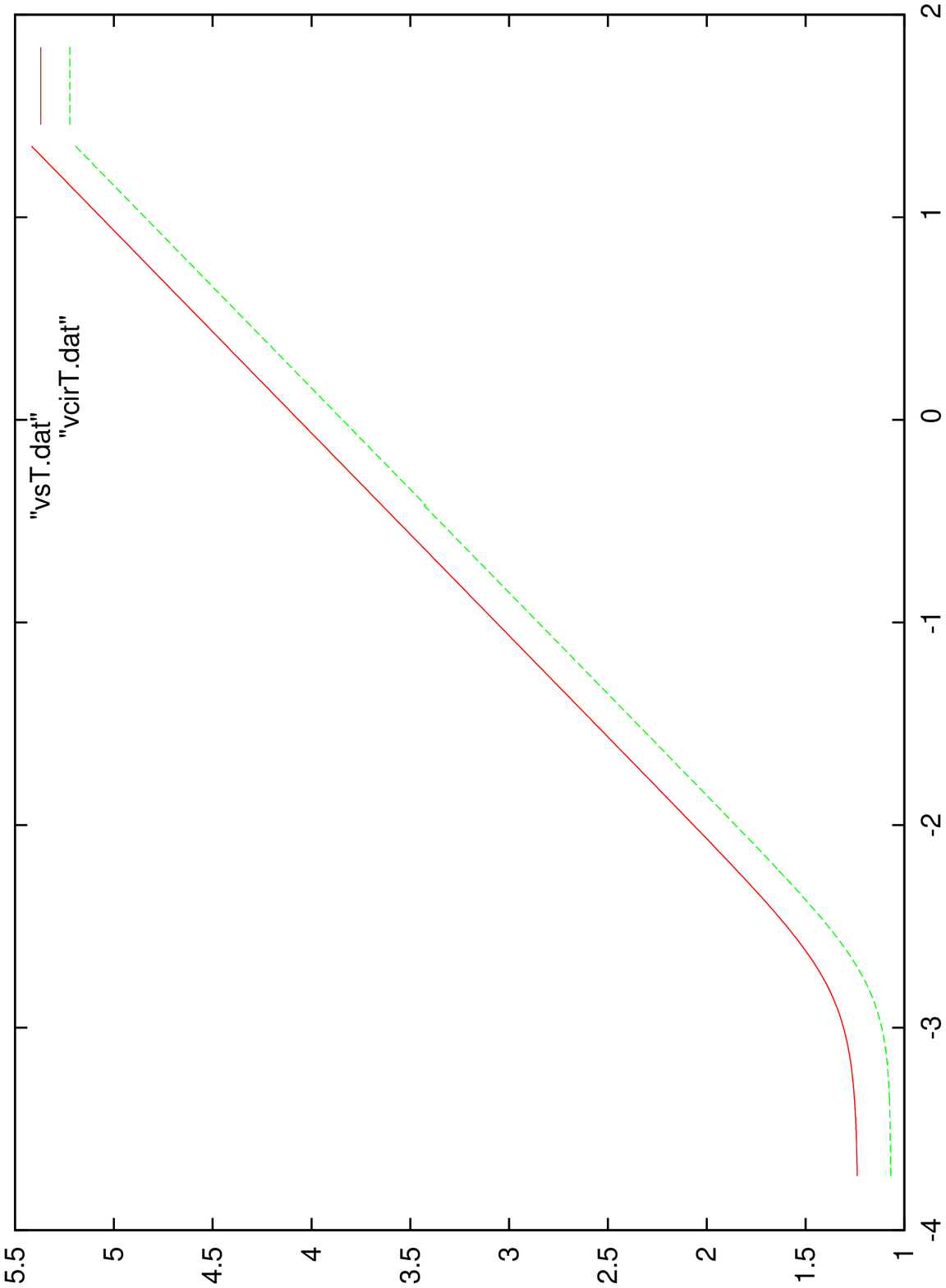}
\psfrag{"vsM.dat"}{$ {\hat v}^2(0) = 3 \; {\hat \sigma}^2(0) $}
\psfrag{"vcirM.dat"}{$ {\hat v}_c^2(r_h) $}
\includegraphics[height=12.cm,width=7.cm]{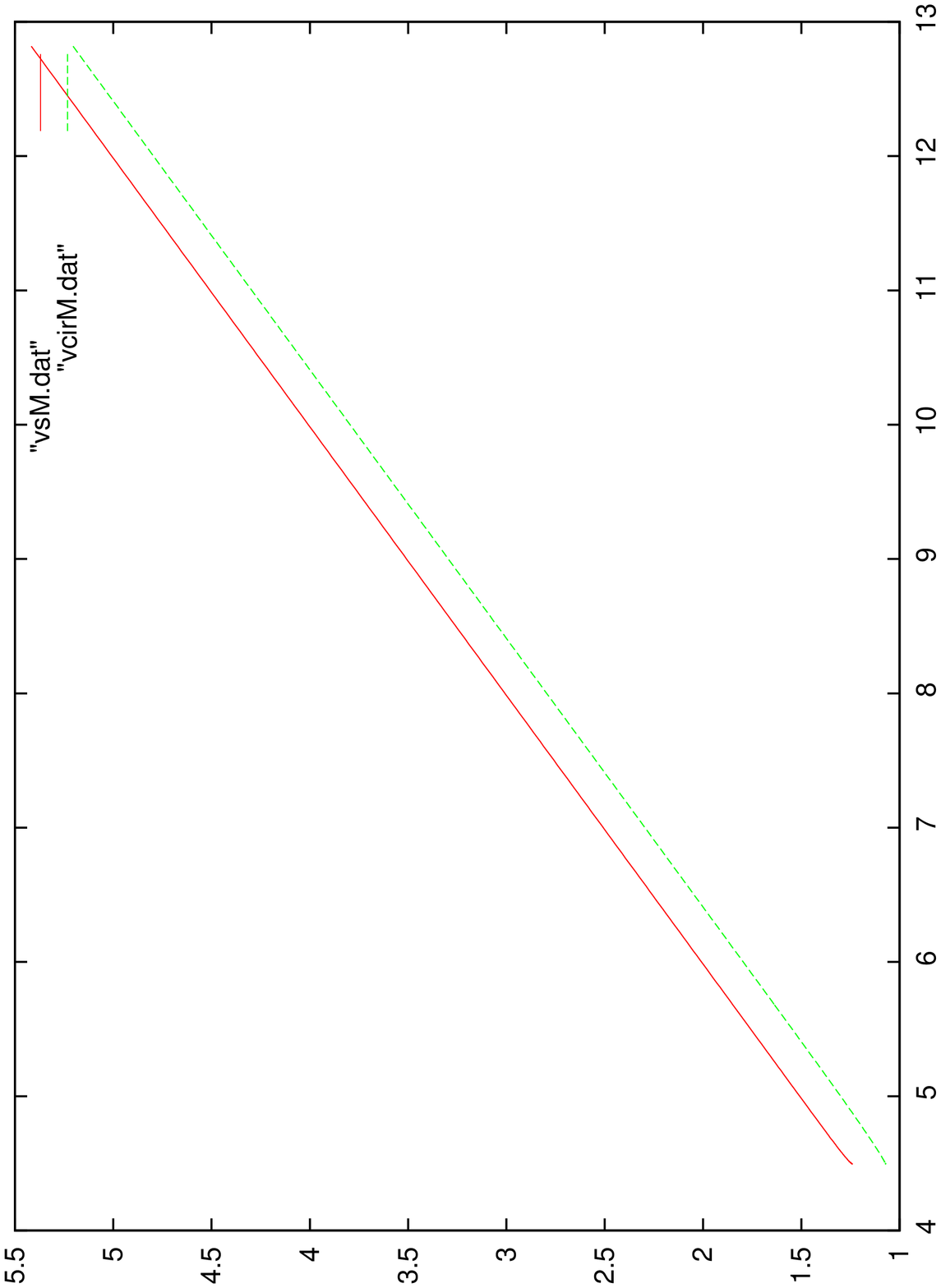}
\end{turn}
\caption{The ordinary logarithm of the velocity dispersion at the origin 
$  {\hat v}^2(0) = 3 \; {\hat \sigma}^2(0) $ and the ordinary logarithm of the circular velocity
at the halo radius $ {\hat v}_c^2(r_h) $ vs. $ \log_{10} {\hat T}_0 $ in the upper panel and vs.
$ \log_{10} {\hat M}_h $ in the lower panel. Notice the unit slope in the upper panel curves as
functions of $ {\hat T}_0 $ according to eq.(\ref{sigdil}) in the diluted regime,
and the one-half slope in the lower panel curves as functions of $ {\hat M}_h $ following 
eqs.(\ref{sigmac})-(\ref{vch}) in the diluted regime. The deviation from the diluted scaling regime
near the degenerate regime is manifest as function of $ {\hat T}_0 $ (upper panel) 
while it is imperceptible as function of $ {\hat M}_h $ (lower panel).}
\label{veloc}
\end{figure}

\begin{figure}
\begin{turn}{-90}
\psfrag{"Kperfu10.dat"}{$ {\hat M}_h= 7 \; 10^{11}  \; M_\odot $}
\psfrag{"Kperfu11.dat"}{$ {\hat M}_h= 6.2 \; 10^8  \; M_\odot $}
\psfrag{"Kperfu12.dat"}{$ {\hat M}_h= 1.3 \; 10^8 \; M_\odot  $}
\psfrag{"Kperfu13.dat"}{$ {\hat M}_h= 2.5 \; 10^7 \; M_\odot  $}
\psfrag{"Kperfu14.dat"}{$ {\hat M}_h= 5.1 \; 10^6 \; M_\odot  $}
\psfrag{"Kperfu15.dat"}{$ {\hat M}_h= 1.1 \; 10^6 \; M_\odot  $}
\psfrag{"Kperfu16.dat"}{$ {\hat M}_h= 2.2 \; 10^5 \; M_\odot  $}
\psfrag{"Kperfu30.dat"}{$ {\hat M}_h= 1.6 \; 10^5 \; M_\odot  $}
\includegraphics[height=12.cm,width=7.cm]{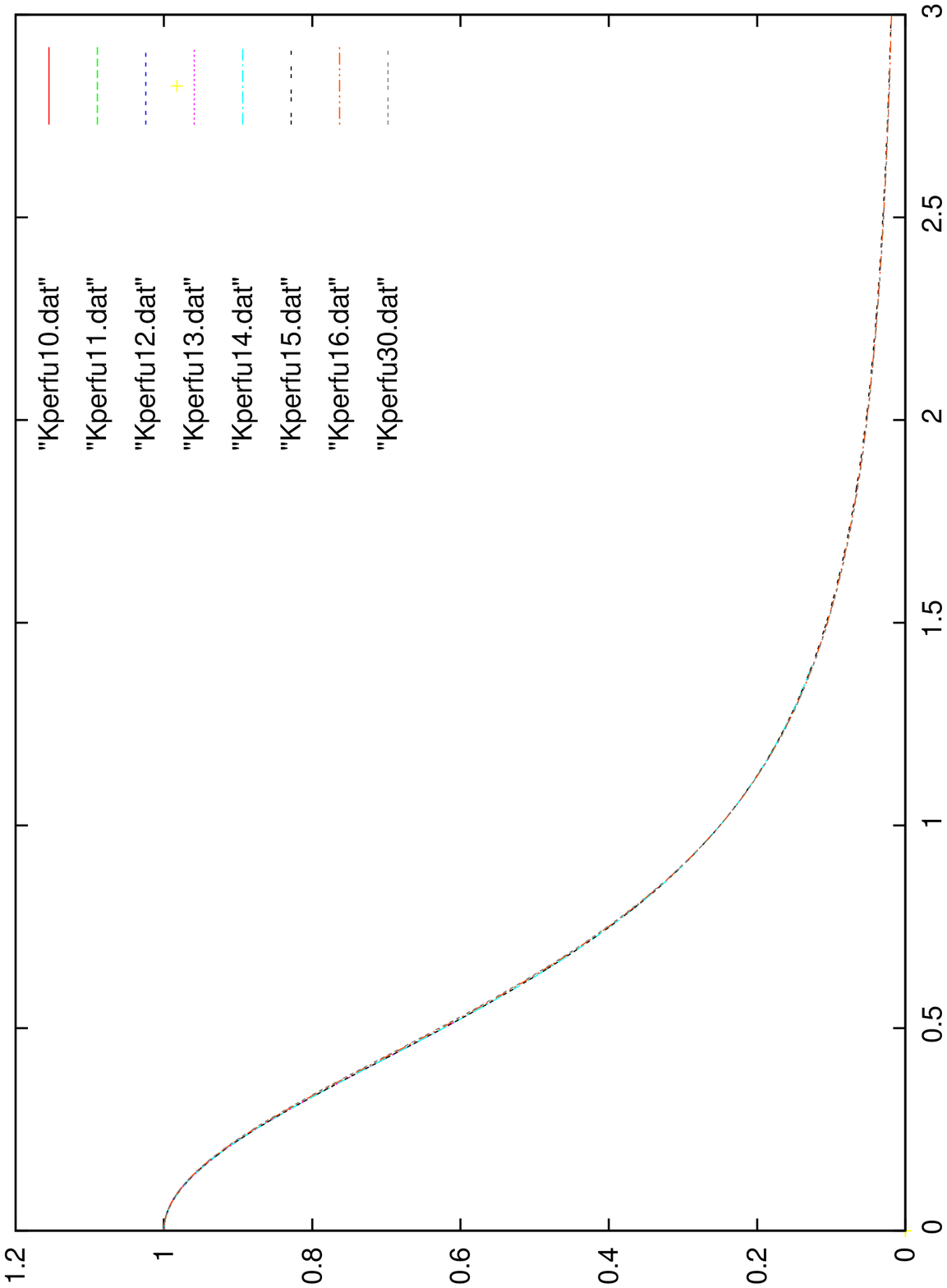}
\psfrag{"Rperfu11.dat"}{universal diluted profile}
\psfrag{"Rperfu17.dat"}{$ {\hat M}_h= 3.9 \; 10^4 \; M_\odot  $}
\psfrag{"Rperfu18.dat"}{$ {\hat M}_h= 2.3 \; 10^4 \; M_\odot  $}
\psfrag{"Rperfu19.dat"}{$ {\hat M}_h= 2.04 \; 10^4 \; M_\odot  $}
\psfrag{"Rperfu20.dat"}{$ {\hat M}_h= 2.0 \; 10^4 \; M_\odot  $}
\psfrag{"Rperfudeg.dat"}{degenerate limit}
\includegraphics[height=12.cm,width=7.cm]{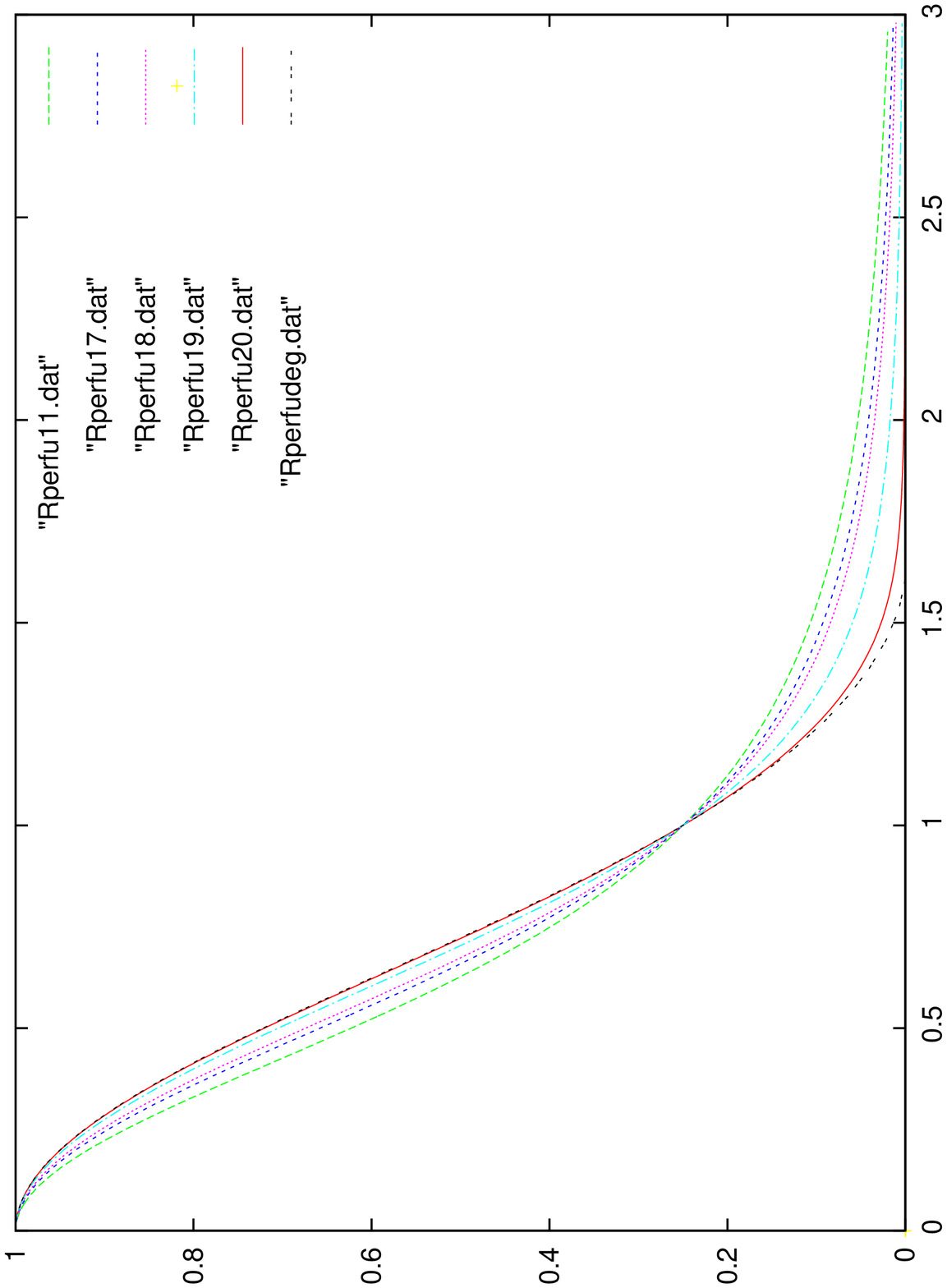}
\end{turn}
\caption{Normalized density profiles $ \rho(r)/\rho(0) $ as functions of $ r/r_h $.
We display in the upper panel the profiles for galaxy masses in the diluted regime
$  1.4 \; 10^5 \; M_\odot < {\hat M}_h < 7.5 \; 10^{11}\; M_\odot , \; -1.5 > \nu_0 > -20.78 $ which {\bf all} provide 
the {\bf same universal} density profile. We display in the lower panel the profiles for
galaxy masses $  M_h^{min} = 30999 \; \left(2 \, {\rm keV}/m\right)^{\! \! \frac{16}5} \; M_\odot
 \leq {\hat M}_h <  3.9 \; 10^4 \; M_\odot, \; 1 < \nu_0 < \infty $ which are near the quantum 
degenerate regime and exhibit shrinking density profiles for decreasing galaxy mass.
For comparison, we also plot in the lower figure the universal profile in the diluted regime.}
\label{perfus}
\end{figure}

\begin{figure}
\begin{turn}{-90}
\psfrag{"Rperfu.dat"}{Exact Thomas-Fermi}
\psfrag{"perf4.dat"}{Empiric formula  $ F_{\alpha=1.5913}(x) $}
{\label{empi} \includegraphics[height=12.cm,width=7.cm]{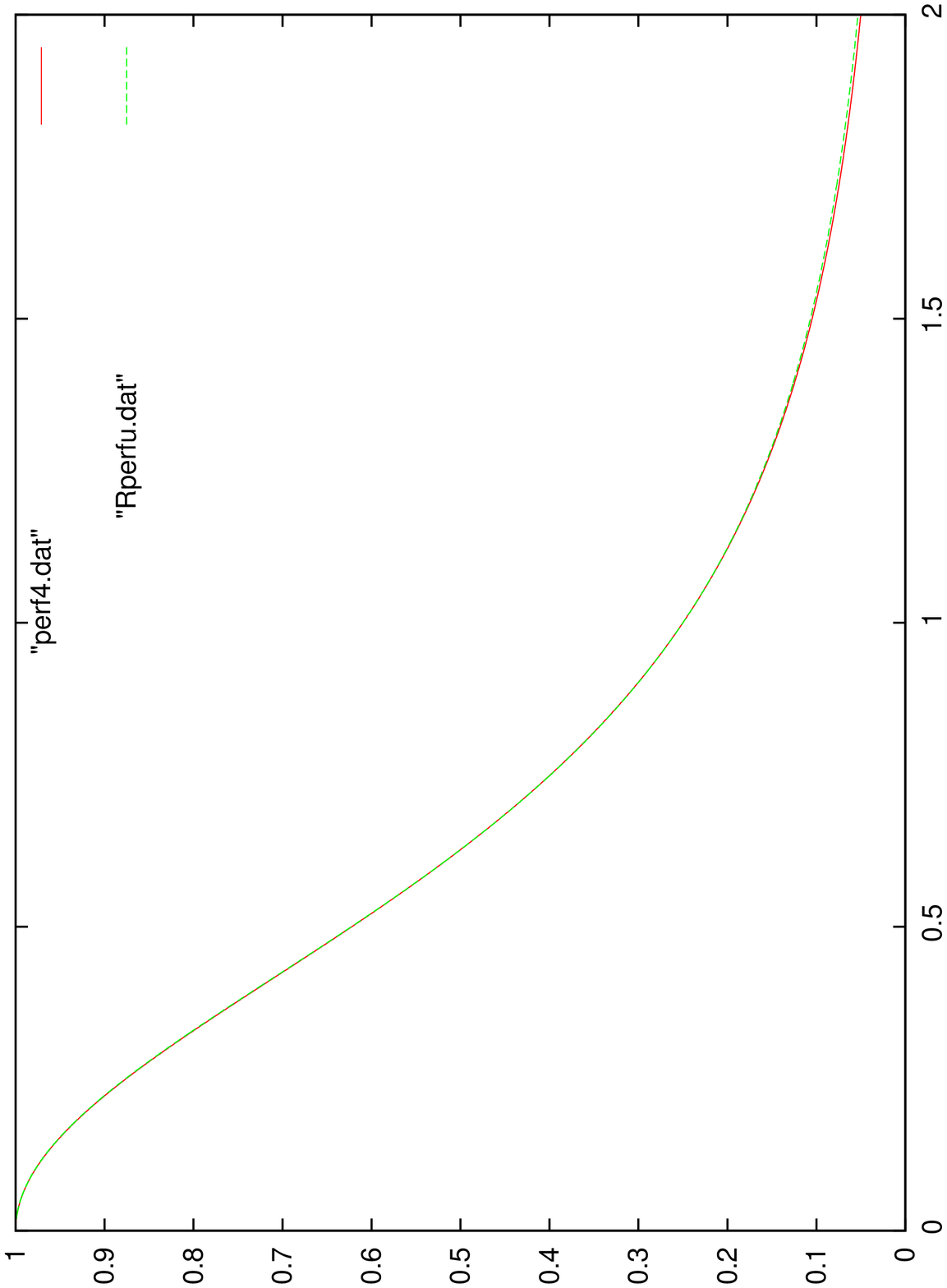}}
\psfrag{"perfu10.dat"}{universal diluted profile}
\psfrag{"asiuniv.dat"}{asymptotic formula}
{\label{asi}\includegraphics[height=12.cm,width=7.cm]{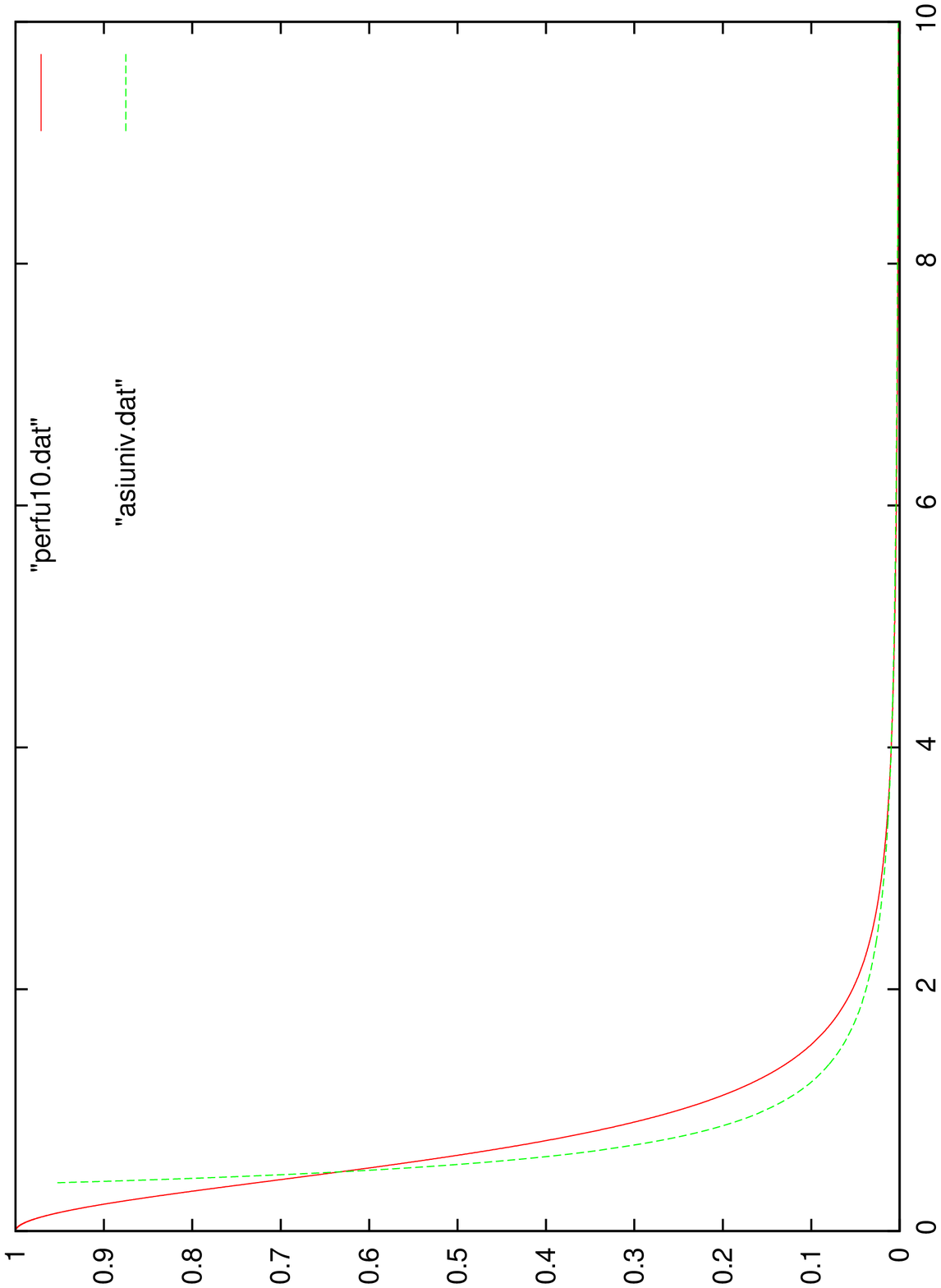}}
\end{turn}
\caption{Upper panel. The universal density profile $ \rho(r)/\rho(0) $ obtained from the Thomas-Fermi
equations plotted vs. $ x = \frac{r}{r_h} $ and its fitting formula (\ref{pempi})
for the best fit value $ \alpha = 1.5913 $. Lower panel.
The universal density profile $ F(x) $ obtained
from the Thomas-Fermi equations vs. $ x = r/r_h $ and its asymptotic form $ F_{asy}(x) $
given by eq.(\ref{asif}). For $ x \gtrsim 3 , \; F_{asy}(x) $ becomes
a very good approximation to $ F(x) $.}
\label{ufa}
\end{figure}

\begin{figure}
\begin{turn}{-90}
\psfrag{"perfv11.dat"}{$ {\hat M}_h= 7.1 \; 10^{11 \; M_\odot } $}
\psfrag{"perfv12.dat"}{$ {\hat M}_h= 6.3 \; 10^9 \; M_\odot  $}
\psfrag{"perfv13.dat"}{$ {\hat M}_h= 1.2 \; 10^8 \; M_\odot  $}
\psfrag{"perfv14.dat"}{$ {\hat M}_h= 2.3 \; 10^6 \; M_\odot  $}
\psfrag{"perfv15.dat"}{$ {\hat M}_h= 2.2 \; 10^5 \; M_\odot  $}
\psfrag{"perfv16.dat"}{$ {\hat M}_h= 6.1 \; 10^4 \; M_\odot  $}
\psfrag{"perfv17.dat"}{$ {\hat M}_h= 4.0 \; 10^4 \; M_\odot  $}
\psfrag{"perfv18.dat"}{$ {\hat M}_h= 3.0 \; 10^4 \; M_\odot  $}
\psfrag{"perfv19.dat"}{$ {\hat M}_h= 2.2 \; 10^4 \; M_\odot  $}
\psfrag{"perfv20.dat"}{$ {\hat M}_h= 2.0 \; 10^4 \; M_\odot  $}
\psfrag{"perfvdeg.dat"}{degenerate limit}
\includegraphics[height=12.cm,width=10.cm]{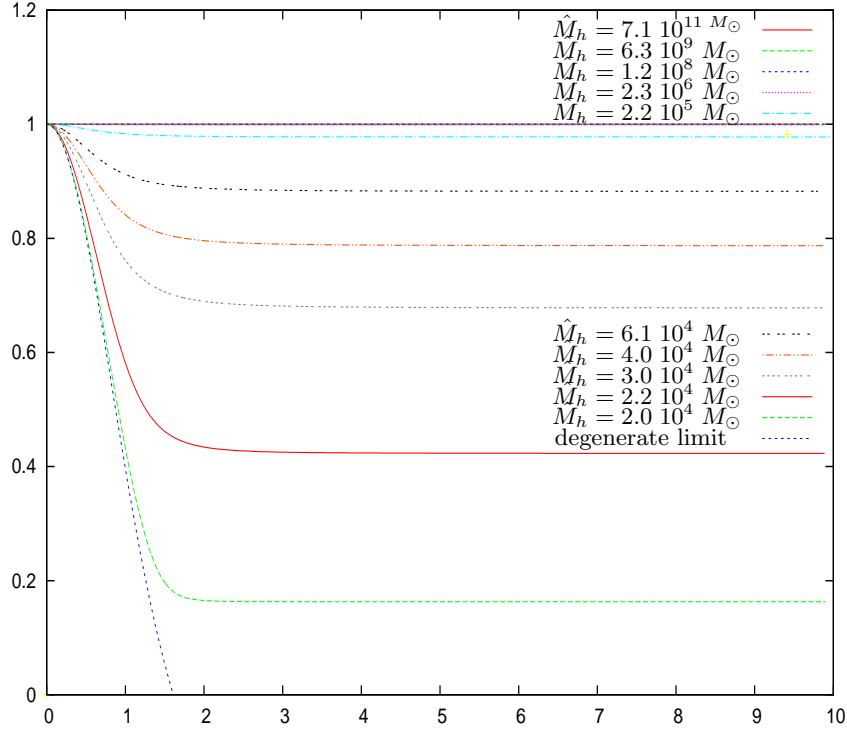}
\end{turn}
\caption{Normalized velocity dispersion profiles $ \sigma^2(r)/\sigma^2(0) $ as functions of $ x=r/r_h $.
All velocity profiles in the diluted regime for galaxy masses $ {\hat M}_h > 2.3 \; 10^6 \; M_\odot, 
\; \nu_0 < -5 $ fall into
the same {\bf constant universal} profile corresponding to a perfect but inhomogeneous self-gravitating WDM
gas describing large and diluted galaxies. 
The velocity profiles for smaller galaxy masses $ 1.6 \; 10^6 \; M_\odot> {\hat M}_h > {\hat M}_{h,min} 
= 3.10 \; 10^4 \; M_\odot $ 
do depend on $ x $ and yield decreasing velocity dispersions for decreasing galaxy masses, accounting for the quantum
fermionic effects which become important in this range of galaxy masses (WDM compact dwarf galaxies). }
\label{perfv}
\end{figure}

\medskip

The opposite limit, $ \nu_0 \gtrsim 1 $, is the quantum regime corresponding to compact
WDM fermions. In particular, in the degenerate limit $ \nu_0 \to \infty $, the galaxy mass and halo radius take
their {\bf minimum } values
\bea\label{rMhmin}
&& r_h^{min} = 11.3794 \; \left(\frac{2 \, {\rm keV}}{m}\right)^{\! \! \frac85}  \; 
\left(\frac{120 \; M_\odot}{\Sigma_0 \;  {\rm pc}^2}\right)^{\! \! \frac15} \; {\rm pc} \; , \cr \cr
&&  M_h^{min} = 30998.7 \; \left(\frac{2 \, {\rm keV}}{m}\right)^{\! \! \frac{16}5} \;
 \left(\frac{\Sigma_0 \;  {\rm pc}^2}{120 \; M_\odot}\right)^{\! \! \frac35}  \;  M_\odot \; ,
\eea
while the phase-space density $ Q(r) $ takes its {\bf maximum} value
\be
Q_h^{max} = 16 \; \frac{\sqrt{125}}{3 \; \pi^2} \;  \left(\frac{m}{2 \, {\rm keV}}\right)^4 \; {\rm keV}^4 =
6.04163 \; \left(\frac{m}{2 \, {\rm keV}}\right)^4 \; {\rm keV}^4 \label{qhdeg} \; .
\ee
From the minimum value of the galaxy mass $ M_h^{min} $ we derive a lower bound for the WDM particle mass
$ m $
\be
m \geq m_{min} \equiv 1.387 \; {\rm keV} \; \left(\frac{10^5 \; M_\odot}{M_h}\right)^{\! \! \frac5{16}} \;
\left(\frac{\Sigma_0 \;  {\rm pc}^2}{120 \; M_\odot}\right)^{\! \! \frac3{16}}
\ee
From the minimal known halo mass $ M_h = 3.9 \; 10^4 \; M_\odot $ for Willman I (see Table 1 in \cite{astro})
we obtain the lower bound
$$
m \geq 1.86 \; {\rm keV} \quad {\rm for ~ Dirac ~ fermions} \quad , \quad
m \geq 2.21 \; {\rm keV} \quad {\rm for ~ Majorana ~ fermions} \; .
$$

\section{Density and velocity dispersion. Universal and non-universal profiles}

It is illuminating to normalize the density profiles as $ \rho(r)/\rho(0) $ 
and plot them as functions of $ r/r_h $. We find that these normalized profiles
are {\bf universal} functions of $ x \equiv r/r_h $ in the diluted regime as shown in fig. \ref{perfus}.
This universality is valid for {\bf all} galaxy masses $ {\hat M}_h > 10^5  \; M_\odot $.

{\vskip 0.1cm} 

No analytic form is available for the profile $ \rho(r) $ obtained from the resolution of the
Thomas-Fermi equations (\ref{nu}). The universal profile $ F(x) = \rho(r)/\rho(0) $ can be fitted
with precision by the simple formula
\be\label{pempi}
F_\alpha(x) = \frac1{\left[1+\left(4^\frac1{\alpha}-1 \right) \; x^2\right]^\alpha} 
\quad , \quad x = \frac{r}{r_h} \quad , \quad \alpha = 1.5913\; .
\ee
The value $ \alpha = 1.5913 $ provides the best fit. We plot in fig. \ref{ufa} $ \rho(r)/\rho(0) $
from the Thomas-Fermi equations (\ref{nu}) and the precise fitting formula $ F_{\alpha=1.5913}(x) $.
The fit is particularly precise for $ r < 2 \; r_h $.

{\vskip 0.1cm} 

Our theoretical density profiles and rotation curves obtained from the 
Thomas-Fermi equations remarkably agree with observations for $ r \lesssim r_h $, for all 
galaxies in the diluted regime \cite{urc}. This indicates that WDM is thermalized in the internal regions 
$ r \lesssim r_h $ of galaxies.

{\vskip 0.1cm} 

The theoretical profile $ \rho(r)/\rho(0) $ and the precise fit $ F_{\alpha=1.5913}(x) $ cannot be used for
$ x \gg 1 $ where they decay as a power $ \simeq 3.2 $ which is a too large number to reproduce the observations.

{\vskip 0.2cm} 

The universal density profile $ \rho(r)/\rho(0) $ is obtained theoretically in the diluted Boltzmann regime.
In such regime the density profile decreases for large $ x \gg 1 $ as $ \sim 1/x^2 $.
More precisely, we find the asymptotic behaviour
\be\label{asif}
F(x) \buildrel_{x \gg 1}\over=  F_{asy}(x) \equiv \frac{0.151869}{x^2} 
\left[ 1 + {\cal O}\left(\frac1{\sqrt{x}}\right)\right] \; .
\ee
We plot in fig. \ref{ufa}  $ F(x) $ and its asymptotic behaviour $ F_{asy}(x) $ vs. $ x $. 
We see that $ F_{asy}(x) $ becomes a very good approximation to $ F(x) $ for $ x \gtrsim 3 $. 
When $ F(x) $ behaves as $ \sim 1/x^2 $ the circular velocity for these theoretical 
density profiles becomes constant as shown in \citep{urc}.

{\vskip 0.2cm} 

For galaxy masses $ {\hat M}_h < 10^5  \; M_\odot $, near the quantum degenerate regime, the normalized density profiles 
$ \rho(r)/\rho(0) $ are not anymore universal and depend on the galaxy mass.

{\vskip 0.1cm} 

As we can see in fig. \ref{perfus} the density profile shape changes fastly
when the galaxy mass decreases only by a factor seven from $ {\hat M}_h = 1.4 \; 10^5  \; M_\odot $ to the minimal galaxy
mass $ {\hat M}_{h,min} = 3.10 \; 10^4 \; M_\odot $. In this narrow range of galaxy masses the density profiles 
shrink from the universal profile till the degenerate profile as shown in fig. \ref{perfus}.
Namely, these dwarf galaxies are more compact than the larger diluted galaxies.

{\vskip 0.2cm} 

We display in fig. \ref{perfv} the normalized velocity dispersion profiles $ \sigma^2(r)/\sigma^2(0) $
as functions of $ x = r/r_h $. Again, we see that these profiles are {\bf universal
and constant}, i. e.  independent of the galaxy mass in the diluted regime for 
$ {\hat M}_h > 2.3 \; 10^6 \; M_\odot , \; \nu_0 < -5 , \; T_0 > 0.017 $ K. The constancy of 
$ \sigma^2(r) = \sigma^2(0) $ in the diluted regime
implies that the equation of state is that of a perfect but inhomogeneous WDM gas. Indeed, from eq.(\ref{equip})
\be\label{sigdil}
\sigma^2(r) = \sigma^2(0) = \frac{T_0}{m} \; ,
\ee
and eq.(\ref{eqest}) implies for the WDM diluted galaxies 
the perfect gas equation of state (\ref{equidea})
where both the pressure $ P(r) $ and the density $ \rho(r) $ depend on the coordinates.

{\vskip 0.1cm} 

For smaller galaxy masses $ 1.6 \; 10^6 \; M_\odot > {\hat M}_h > {\hat M}_{h,min} $,
the velocity profiles do depend on $ r $ and yield decreasing velocity dispersions for decreasing galaxy masses.
Namely, the deviation from the universal curves appears for $ {\hat M}_h < 10^6  \; M_\odot  $
and we see that it precisely arises from the quantum fermionic effects which become important
in such range of galaxy masses.

\section{The equation of state of WDM Galaxies. Classical diluted and compact quantum regimes.}

The WDM galaxy equation of state is by definition the functional relation between the 
pressure $ P $ and the density $ \rho $.

{\vskip 0.2cm} 

From eqs. (\ref{gorda}) and (\ref{pres1}) we obtain separately $ P $ and $ \rho $ at a point $ r $ as
\be\label{eqeg}
\rho = \frac{m^\frac52}{3 \, \pi^2 \; \hbar^3} \; \left(2 \; T_0\right)^\frac32 \; I_2(\nu)
\quad , \quad
P = \frac{m^\frac32}{15 \, \pi^2 \; \hbar^3} \; \left(2 \; T_0\right)^\frac52 \; I_4(\nu) \; .
\ee
These equations express parametrically, through the parameter $ \nu $,
the pressure $ P $ as a function of the density $ \rho $ and therefore provide the WDM galaxy equation of state. 

{\vskip 0.1cm} 

For fermionic WDM in thermal equilibrium $ I_2(\nu) $ and $ I_4(\nu) $ are given as integrals
of the Fermi--Dirac distribution function in eq.(\ref{dfI}). For WDM out of thermal equilibrium eq.(\ref{eqeg})
is always valid but $ I_2(\nu) $ and $ I_4(\nu) $ should be expressed as integrals of the corresponding
out of equilibrium distribution function. In the out of equilibrium case $ T_0 $ is just
the characteristic scale in the out of equilibrium distribution function $ f_{out}(E) = \Psi_{out}(E/T_0) $.
For the relevant galaxy physical magnitudes,
the Fermi--Dirac distribution gives similar results than the 
out of equilibrium distribution functions \citep{astro}.

{\vskip 0.1cm} 

In the two WDM galaxy regimes, classical diluted regime, and degenerate quantum regime, 
we can eliminate $ \nu $ in eqs.(\ref{eqeg}) and obtain
$ P $ as a function of $ \rho $ in close form. Let us take the ratios
$ P/\rho $ and $ P/\rho^\frac53 $ in eqs.(\ref{eqeg}):
\be\label{ratio}
\frac{P}{\rho} = \frac25 \;  \frac{T_0}{m} \; \frac{I_4(\nu)}{I_2(\nu)} \quad , \quad
\frac{P}{\rho^\frac53} = \frac{\hbar^2}5 \; \left(\frac{3 \, \pi^2}{m^4}\right)^{\! \frac23} \; 
\frac{I_4(\nu)}{I_2^\frac53(\nu)}
\; .
\ee
In the diluted limit $ \nu \ll -1 $ we have that
$$
\frac{I_4(\nu)}{I_2(\nu)} \buildrel_{\nu \ll -1}\over= \frac52 
$$
and therefore we obtain for WDM in the diluted limit
the local perfect gas equation of state:
\be\label{equidea}
P(r) =  \frac{T_0}{m} \; \rho(r)  \quad , \quad {\rm WDM ~ diluted ~ galaxies} \; .
\ee
The local perfect WDM gas equation of state eq.(\ref{equidea}) is precisely the equation of state
of the Boltzmann self-gravitating gas \cite{npb}.

{\vskip 0.1cm} 

In the degenerate limit $ \nu \gg 1 $ we have that
$$
\frac{I_4(\nu)}{I_2^\frac53(\nu)} \buildrel_{\nu \gg 1}\over= 1 
$$
and therefore  $ P/\rho^\frac53 $ in eq.(\ref{ratio}) becomes the degenerate fermionic 
equation of state at $ T_0 = 0 $,
\be\label{equdeg}
P = \frac{\hbar^2}5 \; \left(\frac{3 \, \pi^2}{m^4}\right)^{\! \frac23} \; \rho^\frac53 
\quad , \quad {\rm WDM ~ degenerate ~ quantum ~ limit} \; .
\ee

{\vskip 0.2cm} 

\begin{figure}
\begin{turn}{-90}
\psfrag{"Lperho1.dat"}{$ {\hat T}_0 = 10 $ K}
\psfrag{"Lperho3.dat"}{$ {\hat T}_0 = 0.534 $ K}
\psfrag{"Lperho5.dat"}{$ {\hat T}_0 = 0.0285 $ K}
\psfrag{"Lperho7.dat"}{$ {\hat T}_0 = 1.52 \; 10^{-3} $ K}
\psfrag{"Lperho9.dat"}{$ {\hat T}_0 = 0.811 \; 10^{-4} $ K}
\psfrag{"Lperho12.dat"}{$ {\hat T}_0 = 10^{-6} $ K}
\psfrag{"Lperho13.dat"}{$ {\hat T}_0 = 10^{-8} $ K}
\psfrag{"Lperho14.dat"}{$ {\hat T}_0 = 10^{-9} $ K}
\psfrag{"Lprdeg.dat"}{degenerate limit $  {\hat T}_0 = 0 $ K}
\includegraphics[height=12.cm,width=10.cm]{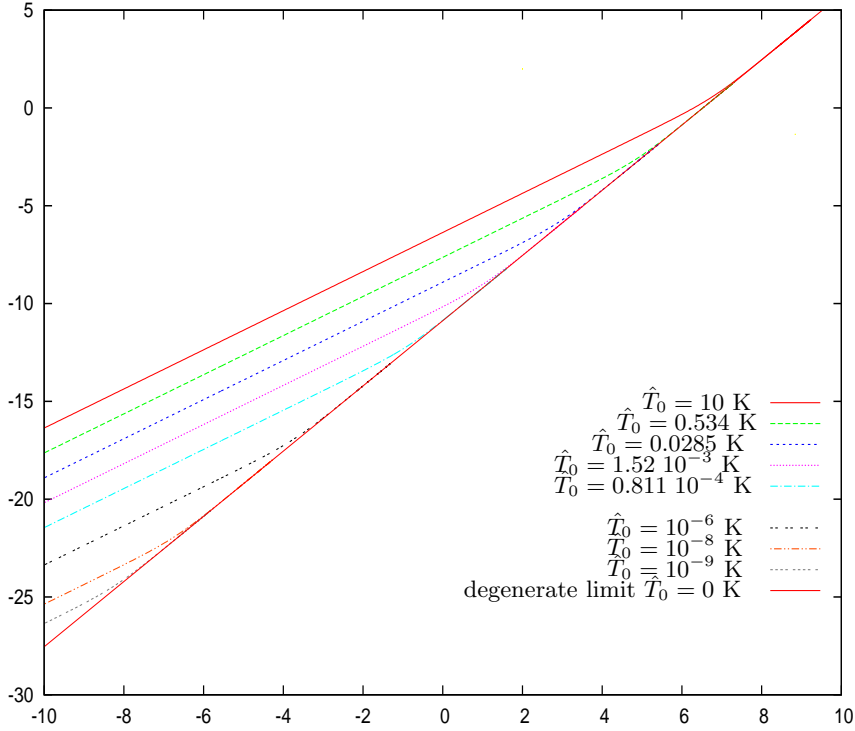}
\end{turn}
\caption{The equation of state of WDM galaxies.
Logarithmic plot of the galaxy pressure $ \bar P $ vs. the density $ \bar \rho $  
as defined by eq.(\ref{barperro}) for different values of the effective temperature
$ T_0 $. For small density and growing $ T_0 $, the self-gravitating ideal WDM gas behaviour is obtained
exhibiting straight lines with unit slope; this describes the physical state of large diluted galaxies
$ {\hat M}_h > 2.3 \; 10^6 \; M_\odot, \; \nu_0 < -5 , \; T_0 > 0.017 $ K.
For large density and decreasing temperature the fermionic 
quantum behaviour close to the degenerate state eq.(\ref{equdeg}) shows up
as the steeper straight lines with slope approaching $ 5/3 $. In particular, the degenerate $ T_0 =0 $
state exhibits the slope $ 5/3 $ for all densities. 
The diluted classical regime and the degenerate regime are interpolated smoothly by the quantum behaviour
corresponding to compact dwarf galaxies with $ 1.6 \; 10^6 \; M_\odot> {\hat M}_h \geq {\hat M}_{h,min} = 
3.10 \; 10^4 \; M_\odot, \; \nu_0 > -4, \; T_0 < 0.011 $ K. For increasing $ T_0 $ the curves move up.
The larger is $ T_0 $, the larger is the value of the density $ {\bar \rho} $ where the
quantum behaviour is attained.}
\label{prho2}
\end{figure}

\begin{figure}
\begin{turn}{-90}
\psfrag{"prho11.dat"}{$ {\hat M}_h= 1.13 \; 10^{12} \; M_\odot  $}
\psfrag{"prho15.dat"}{$ {\hat M}_h= 3.54 \; 10^5 \; M_\odot  $}
\psfrag{"prho16.dat"}{$ {\hat M}_h= 9.72 \; 10^4 \; M_\odot  $}
\psfrag{"prho17.dat"}{$ {\hat M}_h= 6.31 \; 10^4 \; M_\odot  $}
\psfrag{"prho18.dat"}{$ {\hat M}_h= 4.78 \; 10^4 \; M_\odot  $}
\psfrag{"prho19.dat"}{$ {\hat M}_h= 3.48 \; 10^4 \; M_\odot  $}
\psfrag{"prho21.dat"}{$ {\hat M}_h= 4.06 \; 10^4 \; M_\odot  $}
\psfrag{"prho22.dat"}{$ {\hat M}_h= 3.19 \; 10^4 \; M_\odot  $}
\psfrag{"prhodeg.dat"}{degenerate limit $ {\hat M}_h \simeq 3.10 \; 10^4$}
\includegraphics[height=12.cm,width=10.cm]{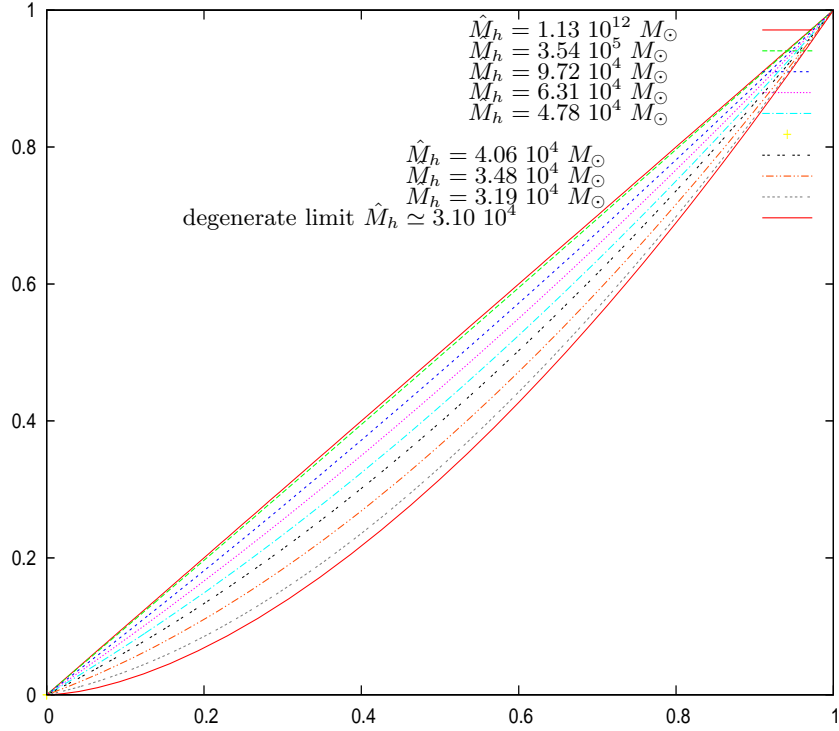}
\end{turn}
\caption{The galaxy pressure $ P/P_0 $ vs. the density $ \rho/\rho_0 $, where $ P_0 $ and
$ \rho_0 $ are the pressure and the density at the origin, respectively defined by eq.(\ref{pyrhonor}).
We see the WDM ideal gas behaviour (unit slope) in the diluted regime, that is for
galaxy masses $ {\hat M}_h > 2.3 \; 10^6 \; M_\odot, \; \nu_0 < -5 , \; T_0 > 0.017 $ K.
For smaller galaxy masses, $ 1.6 \; 10^6 \; M_\odot> {\hat M}_h \geq {\hat M}_{h,min} = 
3.10 \; 10^4 \; M_\odot, \; \nu_0 > -4 $, 
the equation of state depends on the galaxy mass and becomes steeper corresponding to the
quantum fermionic regime of dwarf galaxies. 
In the degenerate limit $ \nu_0 = \infty $ we obtain a 5/3 slope straight line.
We see that the diluted and degenerate regimes are interpolated smoothly by the quantum behaviour.}
\label{prho}
\end{figure}

Making explicit the dimensions, the WDM galaxy equation of state (\ref{eqeg}) becomes
\be
\rho = 4.68591 \; 10^4 \; \left(\frac{T_0}{\rm K}\right)^{\! \! \frac32} \; 
I_2(\nu) \; \left(\frac{m}{2 \, {\rm keV}}\right)^{\! \! \frac52} \; \frac{M_\odot}{{\rm pc}^3} 
\quad , \quad
P = 0.807603 \; 10^{-3} \; \left(\frac{T_0}{\rm K}\right)^{\! \! \frac52} \; I_4(\nu) \;
 \left(\frac{m}{2 \, {\rm keV}}\right)^{\! \! \frac32} \; \frac{M_\odot}{{\rm pc}^3} \; .
\ee
Being the galaxies a nonrelativistic system, $ P $ turns to be much smaller than $ \rho $ 
when both are written in the same units where the speed of light is taken to be unit.

{\vskip 0.1cm} 

It is useful to introduce the rescaled dimensionless variables
\be\label{barperro}
{\bar \rho} \equiv \left(\frac{2 \, {\rm keV}}{m}\right)^{\! \! \frac52} \; \frac{{\rm pc}^3}{M_\odot} \;\rho 
= 4.68591 \; 10^4 \; \left(\frac{T_0}{\rm K}\right)^{\! \! \frac32} \; I_2(\nu)
\quad , \quad
{\bar P} \equiv \left(\frac{2 \, {\rm keV}}{m}\right)^{\! \! \frac32} \; \frac{{\rm pc}^3}{M_\odot} \; P 
= 0.807603 \; 10^{-3} \; \left(\frac{T_0}{\rm K}\right)^{\! \! \frac52} \; I_4(\nu) \; .
\ee
We plot in fig. \ref{prho2} the ordinary logarithm of $ {\bar P} $ vs. the ordinary logarithm of 
$ {\bar \rho} $ for different values of $ T_0 $. For small density and for growing effective temperature,
the self-gravitating ideal WDM gas behaviour eq.(\ref{equidea}) of the diluted regime
is obtained. On the contrary, for large density and for decreasing temperature the fermionic 
quantum behaviour close to the degenerate state eq.(\ref{equdeg}) shows up.
That is, the straight lines with unit slope in fig. \ref{prho2} describe
the perfect WDM gas behaviour eq.(\ref{equidea}), while the steeper straight lines
with slope $ 5/3 $ describe the degenerate quantum behaviour eq.(\ref{equdeg}).
We see that the diluted classical and degenerate regimes are interpolated smoothly by the quantum behaviour.
For increasing $ T_0 $ the curves in fig. \ref{prho2} move up.
The larger is $ T_0 $, the larger is the value of the density $ {\bar \rho} $ where the
quantum behaviour is attained. 

\medskip

We plot in fig. \ref{prho} the pressure normalized to its value at the origin as a function of the
density normalized to its value at the origin according to eqs.(\ref{eqeg}):  
\be \label{pyrhonor}
\frac{P}{P_0} = \frac{I_4(\nu)}{I_4(\nu_0)} \quad {\rm vs.} 
\quad  \frac{\rho}{\rho_0} = \frac{I_2(\nu)}{I_2(\nu_0)} \; .
\ee
The diluted and degenerate gas behaviours eq.(\ref{equidea}) and (\ref{equdeg}) of WDM galaxies are
explicitly seen in fig. \ref{prho}.
The diluted perfect gas behaviour appears for galaxy masses 
$ {\hat M}_h > 2.3 \; 10^6 \; M_\odot  , \; \nu_0 < -5 , \; T_0 > 0.017 $ K.
The degenerate gas behaviour shows up for the minimal mas galaxy 
$ {\hat M}_{h,min} = 3.10 \; 10^4 \; M_\odot  , \; T_0 = 0 $.

{\vskip 0.1cm} 

Besides the two limiting regimes,  diluted and degenerate,
we see from fig. \ref{prho} that the equation of state {\bf does depend} on the galaxy mass
for galaxy masses in the range $ 1.6 \; 10^6 \; M_\odot  > {\hat M}_h \geq {\hat M}_{h,min} ,
\; \nu_0 > -4 , \; T_0 < 0.011 $ K. This is a quantum regime, close to but not at, the degenerate limit.
The equation of state in this quantum regime is steeper than in the degenerate limit. 

{\vskip 0.2cm} 

We find that WDM galaxies exhibit two regimes: classical diluted and quantum compact (close to degenerate).
WDM galaxies are diluted for  $ {\hat M}_h > 2.3 \; 10^6 \; M_\odot  , \; \nu_0 < -5 , \; T_0 > 0.017 $ K
and they are quantum and compact for  
$ 1.6 \; 10^6 \; M_\odot  > {\hat M}_h \geq {\hat M}_{h,min} , 
\; \nu_0 > -4 , \; T_0 < 0.011 $ K. The degenerate limit $ T_0 = 0 $ corresponds to the extreme 
quantum situation yielding a minimal galaxy size $ {\hat r}_{h,min} $ and mass $ {\hat M}_{h,min} $
given by eq.(\ref{rMhmin}).
The equation of state covering all regimes is given by eq. (\ref{eqeg}).

\medskip

We therefore find an explanation for the universal density profiles and universal velocity
profiles in diluted galaxies ($ {\hat M}_h \gtrsim 10^6 \; M_\odot $): these {\bf universal properties}
can be traced back to the perfect gas behaviour of the self-gravitating WDM gas
summarized by the WDM equation of state (\ref{equidea}). Notice that all these universal
theoretical profiles well reproduce the observations for $ r \lesssim r_h $ \citep{urc}.

{\vskip 0.1cm} 

For small galaxy masses, $ 10^6 \; M_\odot  \gtrsim {\hat M}_h \geq {\hat M}_{h,min} = 
3.10 \; 10^4 \; M_\odot  $
which correspond to chemical potentials at the origin $ \nu_0 \gtrsim -5 $ and effective
temperatures $ T_0 \lesssim 0.017 $ K, the equation of state is galaxy mass dependent 
(see fig. \ref{prho}) and the profiles are not anymore universal. These properties account
to the quantum physics of the self-gravitating WDM fermions in the compact case close to the degenerate state.

{\vskip 0.2cm} 

Indeed, it will be extremely interesting to dispose of 
observations which could check these quantum effects in dwarf galaxies.

{\vskip 0.2cm} 

\begin{figure}
\begin{turn}{-90}
\psfrag{"prho.dat"}{Exact Thomas-Fermi}
\psfrag{"2fiteqes.dat"}{Empirical formula}
\includegraphics[height=12.cm,width=10.cm]{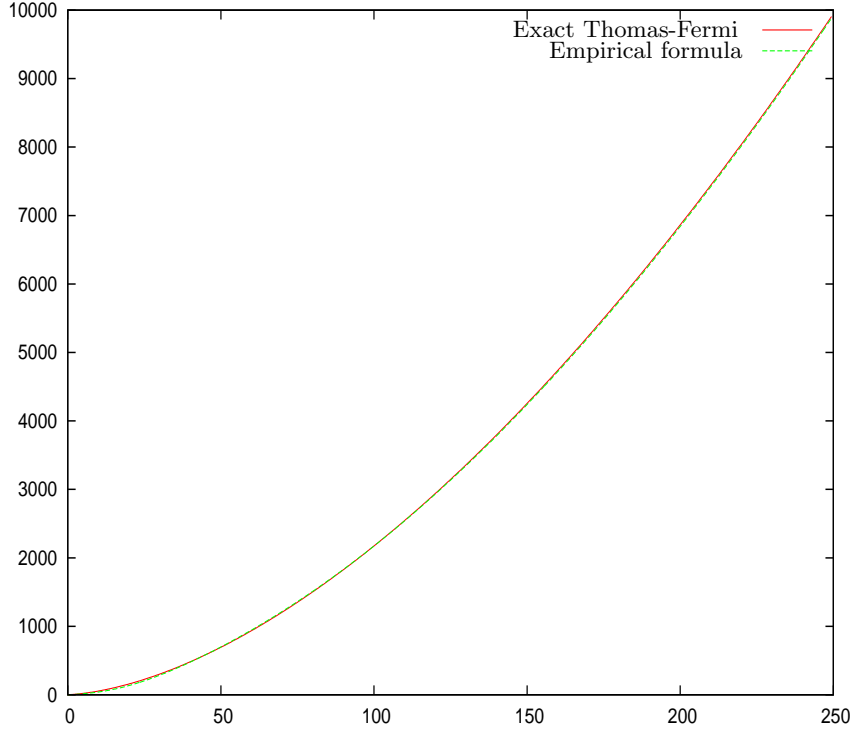}
\end{turn}
\caption{The equation of state $ {\tilde P} $ vs. $ {\tilde \rho} $ obtained by solving
the Thomas-Fermi eq.(\ref{pois}) and the empirical fit eqs.(\ref{defrhot})-(\ref{fitemp}). The exact equation of state
and the fitting formula cannot be distinguished at this resolution.}
\label{eqesfit}
\end{figure}

An useful empiric fit of the exact equation of state follows by expressing the pressure
given by (\ref{eqeg})
\be 
P = \frac{m^\frac32\; \left(2 \; T_0\right)^\frac52 }{15 \, \pi^2 \; \hbar^3} \; {\tilde P}
\quad {\rm where} \quad {\tilde P}  \equiv  I_4(\nu)
\ee
as a function of
\be\label{defrhot}
{\tilde \rho} \equiv  \frac{3 \, \pi^2 \; \hbar^3}{m^\frac52 \; \left(2 \; T_0\right)^\frac32} \; \rho =I_2(\nu)
\ee
We represent $ {\tilde P} $ as a function $ {\tilde \rho} $ through the simple function
\be\label{fitemp}
{\tilde P} = \left(1 + \frac32 \; e^{-\beta_1 \; {\tilde \rho} } \right) \; 
\displaystyle {\tilde \displaystyle \rho}^{\displaystyle\frac13  \displaystyle \left(5 -
2 \;  e^{-\beta_2 \; {\tilde \rho} }\right)}
\ee
that exactly fulfils the diluted and degenerate limiting behaviours (\ref{equidea}) and (\ref{equdeg}), respectively.
Eq.(\ref{fitemp}) best fits the exact values of $ {\tilde P} $ as a function $ {\tilde \rho} $ obtained by solving
the Thomas-Fermi eq.(\ref{pois}) for
\be
\beta_1 = 0.047098 \quad , \quad \beta_2 = 0.064492 \; .
\ee
In summary, we represent the equation of state as
$$
P = \frac{m^\frac32\; \left(2 \; T_0\right)^\frac52 }{15 \, \pi^2 \; \hbar^3} \;\left(1 + 
\frac32 \; e^{-\beta_1 \; {\tilde \rho} } \right) \; \displaystyle 
{\tilde \rho}^{\displaystyle\frac13  \displaystyle \left(5 -
2 \;  e^{-\beta_2 \; {\tilde \rho} }\right)}
$$
where $ {\tilde \rho} $ is expressed in terms of $ \rho $ by eq.(\ref{defrhot}).
We plot in fig. \ref{eqesfit}  $ \; {\tilde P} $ vs. $ {\tilde \rho} $ obtained by solving
the Thomas-Fermi eq.(\ref{pois}) and the empirical fit eq.(\ref{fitemp}).
One can see that the fit turns to be excellent.

\section{The dependence on the WDM particle mass in the diluted and quantum regimes}

In the diluted limit the velocity dispersion is constant $ \sigma^2(r) = \sigma^2(0) $,
eq.(\ref{ecgral}) and eq.(\ref{sigdil}) lead to 
\bea \label{bolpois}
&& \frac{d^2 \mu}{dr^2} + \frac2{r} \; \frac{d \mu}{dr} = - 4\pi \, G \, m \, \rho(r)
\\ \cr
&& \rho(r) = \frac14 \; \left(\frac{2 \, T_0}{\pi \; m}\right)^{\! \! \frac32}
\; m^4 \; \displaystyle \exp\left[\displaystyle\frac{\displaystyle \mu(r)}{T_0}\right] 
\label{bario} \; .
\eea
In this diluted limit, the Thomas-Fermi equations (\ref{pois}) become the
equations for a self-gravitating Boltzmann gas in thermal equilibrium.

Eq.(\ref{bario}) combined with the chemical potential expression (\ref{potq})
becomes the baryotropic equation
$$
\rho(r) = \rho_0 \; e^{- \displaystyle\frac{m}{T_0} \;\left[\phi(r)-\phi(0)\right] } \; .
$$

{\vskip 0.2cm} 

It is instructive to discuss from eq.(\ref{bolpois}) the dependence on the mass $ m $
of the WDM particle. 

{\vskip 0.1cm} 

In the diluted regime,  $ T_0 $ and $ \mu(r) $ depend on $ m $, while other 
magnitudes as $ \rho(r), \; M(r), \;  \sigma^2(r), \; P(r), \; Q(r) $ and $ \phi(r) $
do not depend on $ m $. This means that a change in $ m $, namely
$$
m \Rightarrow m' 
$$
must leave  eq.(\ref{bolpois}) invariant, which implies
$$
\frac{T_0}{m} = {\rm invariant} \quad , \quad  
m^4 \; \displaystyle \exp\left[\displaystyle\frac{\displaystyle \mu(r)}{T_0}\right] = {\rm invariant} \; .
$$
That is,
\be\label{trafm}
T_0(m') = \frac{m'}{m} \; T_0(m) \quad , \quad  
\mu(m',r) = \frac{m'}{m} \left[\mu(m,r) + 4 \;  T_0(m) \; \ln \left(\frac{m}{m'} \right)\right] \; .
\ee
A change in the WDM particle mass $ m $ implies that the temperature $ T_0 $ and the
chemical potential $ \mu(r) $ transform as given by eq.(\ref{trafm}). These transformations
leave the Boltzmann gas equations (\ref{bolpois})-(\ref{bario}) invariant.

{\vskip 0.1cm} 

Under changes of $ m $ the dimensionless variables $ \xi $ and $ \nu(\xi) $ transform as
\be\label{traf2}
m \Rightarrow m' \quad , \quad  
\xi' = \xi \;  \left(\frac{m'}{m}\right)^2 \quad , \quad  \nu(\xi',m') = \nu(\xi,m) + 
4 \;  \ln \left(\frac{m}{m'} \right) \; .
\ee
We see that all the diluted regime relations eqs.(\ref{dilu})-(\ref{vcird}) are invariant under the change 
$ m \Rightarrow m' $ implemented through eqs.(\ref{trafm})-(\ref{traf2}).

\medskip

Indeed, this invariance is restricted to the diluted regime ($ {\hat M}_h \gtrsim 10^6 \; M_\odot $). 

\medskip

For galaxy masses $  {\hat M}_h < 10^5 \; M_\odot $, namely in the {\bf quantum regime of compact dwarf galaxies},
all physical quantities {\bf do depend} on the DM particle mass $ m $ as explicitly displayed
in eqs.(\ref{gorda})-(\ref{vtf}). It is precisely this dependence on $ m $
that leads to the lower bound $ m> 1.91 $ keV from the minimum observed galaxy mass \cite{astro}.
Moreover, for $ m > 2 $ keV, an overabundance of small structures appear as solution of the Thomas-Fermi 
equations, which do not have observed counterpart. Therefore, $ m $ between 2 keV and 3 keV is 
singled out as the most plausible value \cite{astro}.

\medskip

In summary, we see the power of the WDM Thomas-Fermi approach to describe
the structure and the physical state of galaxies
in a clear way and in very good agreement with observations.

\end{document}